\documentclass[12pt]{article}
\usepackage{amssymb,amsmath}
\makeatletter
\usepackage{appendix}
\usepackage{float}

\@addtoreset{equation}{section}
\def\section{\@startsection {section}{1}{\z@}{-2.1ex plus -1ex minus
 -.2ex}{1.3ex plus .2ex}{\large\bf}}
\def\subsection{\@startsection{subsection}{2}{\z@}{-2.0ex plus%
 -1ex minus -.2ex}{0.4ex plus .2ex}{\bf}}

\advance \voffset by -0.9in
\advance \hoffset by -0.7in
\def\cdott{\! \cdot \!}

\def\calP{{\mathcal P}}

\def\be{{\mbox{\boldmath $e$}}}

\def\bn{\mathbf{n}}
\def\bm{\mathbf{m}}

\def\bpm{\begin{pmatrix}}
\def\epm{\end{pmatrix}}

\newcommand{\RR}{\mathbb{R}}
\newcommand{\CC}{\mathbb{C}}

\newcommand{\HH}{\mathbb{H}}

\def\bea{\begin{eqnarray}}
\def\eea{\end{eqnarray}}

\newcommand{\C}{\mbox{${\mathbb C}$}}
\newcommand{\R}{\mbox{${\mathbb R}$}}

\newcommand{\bv}[1]{\mbox{$\bf #1$}}

\newcommand{\CP}{\hbox{{$\mathcal P$}}}

\newcommand{\cm}{\mathfrak{m}}  
\newcommand{\cg}{\mathfrak{g}}

\newcommand{\ch}{\mathfrak{h}}  

\newcommand{\PT}{\tilde{P}}

\newcommand{\tens}{\mathop{\otimes}}
\newcommand{\la}{{\triangleright}}
\newcommand{\ra}{{\triangleleft}}

\def\rcross{{\triangleright\!\!\!<}}
\def\lcross{{>\!\!\!\triangleleft}}

\def\rlbicross{{\triangleright\!\!\!\blacktriangleleft}}
\def\lrbicross{{\blacktriangleright\!\!\!\triangleleft}}
\def\dcross{{\bowtie}}

\def\rbiprod{{\cdot\kern-.33em\triangleright\!\!\!<}}
\def\lbiprod{{>\!\!\!\triangleleft\kern-.33em\cdot}}

\begin{document}
\parskip 4pt
\parindent 8pt
\begin{flushright}
EMPG-11-23
\end{flushright}

\begin{center}

{\Large \bf  On  the semiduals of local isometry groups  in 3d gravity}

\baselineskip 20 pt

\vspace{.2cm}

{ \bf Prince~K.~Osei } \\
Department of Mathematics  \\
 University of Ghana, 
PO Box LG 25,
Legon, Ghana \\
{pkosei@ug.edu.gh}

\vspace{.2cm}

 {\bf  Bernd~J.~Schroers}   \\
Department of Mathematics and Maxwell Institute for Mathematical Sciences \\
 Heriot-Watt University, 
Edinburgh EH14 4AS, United Kingdom \\ 
{b.j.schroers@hw.ac.uk}

\vspace{0.3cm}

{May  2012}
\baselineskip 16 pt

\end{center}

\begin{abstract}
\noindent
We  use factorisations of the local isometry groups arising in 3d gravity for Lorentzian and Euclidean signatures and any value of the  cosmological constant  to  construct associated bicrossproduct quantum groups via semidualisation. In this way we  obtain quantum doubles of  the Lorentz and rotation groups in 3d,  as well as $\kappa$-Poincar\'e algebras whose associated $r$-matrices have  spacelike, timelike and lightlike deformation parameters.  We  confirm and elaborate the  interpretation of semiduality   proposed in \cite{MajidSchroers}  as the exchange of  the cosmological length scale   and  the Planck mass in the context of 3d quantum gravity.  
In particular, semiduality gives a simple understanding of why the quantum double of the Lorentz group and the  $\kappa$-Poincar\'e algebra with spacelike deformation parameter are both associated to 3d gravity with vanishing cosmological constant, while the $\kappa$-Poincar\'e algebra with a timelike deformation parameter can only be associated to 3d gravity if one takes the  Planck mass to be imaginary. 
\end{abstract}


\section{Introduction }
\subsection{Background: 3d quantum gravity and semiduality} 
In three dimensions (two space and one time), every solution  of the Einstein equations  is  locally isometric to a model spacetime which is determined by the signature and the sign (or vanishing) of the cosmological constant \cite{Carlipbook}.  The isometry groups  of the local model spacetimes play a fundamental role in 3d  gravity: elements of the isometry group provide the  gluing data with which globally non-trivial solutions of the Einstein equations on a general 3-manifold are  constructed from copies of the model spacetimes; in  the Chern-Simons formulation of 3d gravity \cite{AT,Witten}, the local isometry groups play the role of gauge groups. 

The space of all gluing data, suitably defined, constitutes the phase space of 3d gravity. It depends on three physical constants: the squared  speed of light $c^2$, the cosmological constant $\Lambda_C$  (both of which affect the structure constants of the isometry Lie algebras)  and the gravitational constant $G$ (which affects the Poisson structure). In this paper we think of the signs or vanishing of these constants  as charactersing different classical regimes of 3d gravity. We  distinguish Lorentzian  and Euclidean regimes  by the sign of $c^2$ (we do not consider the Galilean limit here, but see \cite{PapageorgiouSchroers1}), regimes with positive, negative or vanishing cosmological constant, and finally regimes where the gravitational interaction is switched on or  off.

 Quantisation may be  viewed as  deformation of this classical   picture
into a non-commutative setting, with model spacetimes being replaced by non-commutative spaces and the local  isometry groups by quantum groups. 
This point of view is summarised in  \cite{SchroersCracow}, and based on 
 the application of the combinatorial or Hamiltonian quantisation programme to  the Chern-Simons formulation of 3d gravity \cite{AGSI,AGSII,AS,Schroers,BNR,MeusburgerSchroers1, MeusburgerSchroers2}.
The quantum group deformation of  a given  classical isometry group, called quantum isometry group
 in the following, is found via a classical $r$-matrix which is required to be compatible with the Cherns-Simons action in a certain sense \cite{SchroersCracow}.

The combinatorial quantisation procedure does not define the quantum isometry group
uniquely. Instead, it defines an equivalence class of quantum groups,   with equivalence essentially given by  twisting. 
 In the combinatorial approach this ambiguity does not matter since constraints are imposed after quantisation, and   equivalent quantum groups give rise to the same quantum theory once the constraints are imposed.  Thus, one may view combinatorial quantisation as a map from a classical regime and a given spacetime topology to a quantum theory of 3d gravity.

 It is conceivable that there are equivalences between quantum groups which  mean that the  same quantum theory is associated  to   classically distinct regimes. In the interpretation of  our results (though not in their derivation) we  shall assume that this is not the case, i.e. we assume a one-to-one correspondence between   regimes of 3d gravity and equivalence classes of quantum isometry groups. This allows us to read off the regime from a given  quantum group.  At the end of this introduction we shall comment on ways of establishing the validity this assumption.

An additional feature of the general picture, observed and discussed in \cite{MajidSchroers} in the Euclidean setting, is that  several of the quantum groups in the family of quantum doubles and bicrossproduct quantum groups associated to  3d gravity  are  related by a map called semidualisation. As elaborated in \cite{MajidSchroers}, this map, which is sometimes referred to as Born reciprocity,  can be interpreted in two ways.

The first way is to think of it as an exchange  of position and momentum degrees of freedom.  For example, applying  semiduality to the universal enveloping algebra of the Euclidean Lie algebra $\mathfrak{so}(3)\ltimes \R^3$,  one keeps the angular momentum generators (hence the prefix semi-)   and  replaces   momenta (which generate translations in space) by  position coordinates (which generate translations in momentum space). In this simple example, the resulting   algebra  is isomorphic to the original one, essentially because position and momentum space are isometric in this case. 
  
  Another  interpretation of semiduality, and the one that is most interesting in 3d gravity, is to interpret both the original and the semidual generators in the same way, but to think of semiduality as a map between different regimes. An example studied in \cite{MajidSchroers} is the universal enveloping algebra $U(\mathfrak{so}(4))$, which semidualises to the quantum double $D(U(\mathfrak{su}(2)))$. In 3d gravity, the former is associated with a positive cosmological constant and vanishing gravitational constant, whereas the latter  is associated with vanishing cosmological constant and positive gravitational constant.
  
\subsection{Motivation and outline} 
The purpose of the current paper is to explore further  the interpretation and role  of semi\-duality in 3d gravity  and to extend the discussion to  the Lorentzian setting. We show that the  quantum doubles  of $U(\mathfrak{su}(2))$ and $U(\mathfrak{sl}(2,\RR))$ as well as  the three-dimensional $\kappa$-Euclidean and $\kappa$-Poincar\'e algebras with timelike, lightlike and spacelike deformation parameters can all be obtained as semi-duals of  the  universal enveloping algebras of  the  local isometry Lie algebras of   3d gravity.  Our calculations     generalise and unify  those carried out in \cite{MajidRuegg} in the construction of the $\kappa$-Poincar\'e algebra in 3+1 dimensions \cite{LNRT,LNR} as a bicrossproduct  and   those in \cite{Majidbicrossexample,Majidbook1}  in the  3d Euclidean setting. 

Computing the semidual of a Hopf algebra requires that one chooses a factorisation of the original algebra as a double cross product. We consider the universal enveloping algebras of  each of the  isometry Lie algebras, but 
do not consider all possible factorisations here, leaving  a systematic discussion for  future work \cite{OseiSchroers3}. Instead, we focus on those factorisations which lead to semiduals which are either quantum doubles or $\kappa$-Euclidean and $\kappa$-Poincar\'e algebras.  These are  the quantum groups whose relationships and roles  in the context of  3d gravity we would like to understand. 

 It turns out that the algebra structure of the semidual Hopf algebra only depends on the signature of spacetime: it is the  universal enveloping algebra of the Euclidean Lie algebra in the Euclidean case,  and the universal enveloping algebra of the Poincar\'e Lie algebra in the Lorentzian case. However, the co-algebra structure depends on the signature, the cosmological constant and also on the chosen factorisation. 

We also consider the Lie bialgebras associated to each of the  semidual Hopf algebras,  and   compute  their classical $r$-matrices. Thus, at the end of a long chain of calculations we obtain a map from  an isometry group together with a chosen factorisation  to a classical $r$-matrix for either the Euclidean or Poincar\'e Lie algebra. If the factorisation does not depend on a vector, the $r$-matrix is always that associated to the classical double structure of the Euclidean or Poincar\'e Lie algebra in 3d. If the factorisation does depend on a vector, the $r$-matrix  also depends on the vector and is of bicrossproduct type. 

 At the end of our paper, we summarise our results by  interpreting semiduality as a map between regimes, as explained above. In the cases considered in this paper, the semidual regime always has a vanishing cosmological constant (as indicated by the algebra structure) but it may have vanishing, real or possibly imaginary gravitational constant.  More precisely, it turns out that the cosmological curvature radius in the original theory (suitably interpreted in the Lorentzian case) becomes the Planck mass (which is essentially the inverse gravitational constant in 3d gravity)  in the semidual theory.

Our results also suggest an interesting new viewpoint for understanding the ambiguity in determining the quantum isometry group in 3d gravity.
We observe  that, in the examples considered in this paper,  the $r$-matrices computed for semiduals of different factorisation of the same algebra are  twist-equivalent. In the context of the combinatorial quantisation programme this means that  those  semiduals of a given isometry algebra (but for different factorisation) are  equally valid as quantum isometry groups in the semidual regime.  Thus we see that, in 3d gravity and for the examples considered here,  twist-equivalent quantum isometry groups in the semidual regime can be understood as  simply arising from different factorisations of {\em one} given isometry algebra in the original regime.  This suggests that one may, more generally,   obtain a better understanding of the twist-equivalence of quantum isometry groups by studying their semiduals.


For a full and general justification of our interpretation of semiduality as a map between physical regimes it is important to clarify if  semiduality relations between quantum groups descend to semiduality relations between   equivalence classes of quantum groups  in the combinatorial approach to 3d quantum gravity. In order to do this one first needs  to   characterise (or explicitly list)  the equivalence classes of quantum isometry groups in the combinatorial approach to 3d gravity. This requires, in turn, the classification of  classical $r$-matrices which are compatible with a given Chern-Simons action.  There are partial results on the classification of compatible classical $r$-matrices in \cite{Stachura} in the case of vanishing cosmological constant,  and we give  a full classification of the possible $r$-matrices compatible with a generalised Chern-Simons action for all signatures and values of the cosmological constant in a forthcoming paper \cite{OseiSchroers2}. However, finding all associated quantum groups presents a further and considerable challenge. 

Once the equivalence classes are understood one can check if semiduality respects the equivalence. 
  Our examples  where  different factorisations of  one algebra give rise to twist-equivalent quantum groups suggests that this may be the case. However, many more examples or, even better, a general argument  are required to  clarify the picture.

This paper is organised  as follows. In Sect.~2 we review the local isometry groups arising in 3d gravity and describe the associated Lie algebras in three different ways: via their Cartan decomposition, via their Iwasawa decomposition and, in a unified language for all values of the cosmological constant and both signatures, in terms of (pseudo) quaternions. Sect.~3 contains a short summary of relevant facts about   bicrossproduct Hopf algebras and the mathematical process of semidualisation. In Sect.~4 we use the Iwasawa decomposition of the local isometry groups and the unified quaternionic language developed in \cite{MeusburgerSchroers6} to exhibit double cross product structures of  the local isometry groups. By definition, double cross product  groups can be factorised into two subgroups. Factorising in different orders gives rise to  actions of the two factors on each others. These  transformations are key in constructing the semidual Hopf algebras  and we therefore review them in Sect.~4 as well. In Sect.~5 we carry out the semidualisation  for each chosen  factorisation of the local isometry groups, and construct the associated bicrossproduct Hopf algebras. In our final Sect.~6 we show that the Lie bi-algebras for each of the bicrossproducts constructed via semidualisation are  co-boundary, compute the associated classical $r$-matrices in each case and discuss  our results in the  general context  summarised in this introduction.  The appendix contains the general results of Sect.~5 in more familiar and standard notation for each signature  and  sign (or vanishing) of the cosmological constant.

\section{Local isometry groups of 3d gravity and their Lie algebras}

\subsection{Conventions}
The notation and concepts introduced in this section closely follow \cite{MeusburgerSchroers6}.
We use Einstein's summation convention  and raise 
indices with either the three-dimensional Euclidean
metric $\text{diag}\left(1,1,1\right)$ or the three-dimensional
Minkowski metric $\text{diag}(1,-1,-1)$. A convenient way of unifying  the Euclidean and Lorentzian view point is to write 
\begin{equation}
\eta=\text{diag}\left(1,-\frac {|c|^2}{c^2}, -\frac{|c|^2}{c^2}\right).
\end{equation}
Then we recover the Lorentzian metric   for real  speed of light $c$ and the Euclidean metric  for imaginary  $c$. 
We do not introduce separate notation for the Lorentzian and the Euclidean case since the distinction will be clear from the context. 
In particular,  we write, in either case,
\begin{equation}
\textbf{p}\cdott\textbf{q}=\eta_{ab}p^{a}q^{b},\mbox{ with }\textbf{p}=(p^{0},p^{1},p^{2}),\textbf{q}=(q^{0},q^{1},q^{2})\in\mathbb{R}^{3}.
\end{equation}

The generators of both the three dimensional rotation
algebra  $\mathfrak{su}(2)$ and the three-dimensional Lorentz
algebra $\mathfrak{sl}(2,\R)$ are denoted by $J_{a}$, with the distinction between the two cases again given by the context.
 In terms of these generators the Lie brackets
and Killing form are respectively 
\begin{equation}
\label{basicbracket}
\left[J_{a},J_{b}\right]=\epsilon_{abc}J^{c}\quad \mbox{ and }\quad  \kappa\left(J_{a},J_{b}\right)=\eta_{ab},\end{equation}
where $\epsilon$ denotes the fully antisymmetric tensor in three
dimensions with the convention $\epsilon_{012}=\epsilon^{012}=1$.  We write $\ch$ for the Lie algebra with brackets  \eqref{basicbracket} if we do not need to distinguish between the Lorentzian and Euclidean case.

\subsection{Cartan decomposition}
The solutions of the Einstein equations in three dimensions are locally isometric to certain  model spacetimes which are completely 
determined by the signature of spacetime and the cosmological constant. The isometry groups of these model spacetimes are local 
isometries of 3d gravity. The corresponding Lie algebras can be expressed
in a common form in which the cosmological constant $\Lambda_C$ and the speed of light $c$ appear as a parameter in the Lie bracket. 
To achieve this, we define
 \begin{equation}
 \label{lambdadef}
\lambda= -c^2\Lambda_C,
\end{equation}
so that  with $c=1$ in the Lorentzian case  we have  $\lambda =-\Lambda_C$,  and with $c=i$ in the Euclidean case we have  $\lambda =\Lambda_C$. Note that the constant $\lambda$ was denoted $\Lambda$ in \cite{MeusburgerSchroers6}. We attach the subscript $C$ to  the cosmological constant to distinguish it from the constant $\Lambda$ in that paper.

We write $\cg_{\lambda}$ for  the family of  Lie algebras arising in 3d gravity. 
A basis with  a clear physical interpretation  is the Cartan basis, consisting  of generators
$J_{a}$ and $P_{a},$ $a=0,1,2.$ The $J_a$ are the generators of the Lorentz group, where $J_0$ is the rotation generator, $J_1$ and $J_2$ are the boost generators and $P_a$ are the translation generators, with Lie brackets
\begin{equation}
\left[J_{a},J_{b}\right]=\epsilon_{abc}J^{c},\;\left[J_{a},P_{b}\right]=\epsilon_{abc}P^{c}\;\mbox{ and }\;\left[P_{a},P_{b}\right]=\lambda \epsilon_{abc}J^{c}.
\label{eq:lie algebra}
\end{equation}
For $\lambda=0$, the bracket of the generators $P_{a}$ vanishes
and the Lie algebra $\cg_{\lambda}$ is the three-dimensional Euclidean
or Poincar\'e  Lie algebra. For $\lambda<0$, the Lie brackets (\ref{eq:lie algebra}) are those of 
$\mathfrak{so}(3,1)\simeq \mathfrak{sl}\left(2,\mathbb{C}\right)_{ \mathbb{R}}$. They can be  obtained via the identification $P_{a}=i\sqrt{\left|\lambda\right|}J_{a}$
as the complexification of  $\mathfrak{su}\left(2\right)$
and  $\mathfrak{sl}\left(2,\mathbb{R}\right)$ for Euclidean
and Lorentzian signature, respectively. For $\lambda>0$, the alternative
 generators \begin{equation}                          
J_{a}^{\pm}=\frac{1}{2}\left(J_{a}\pm\frac{1}{\sqrt{\lambda}}P_{a}\right)\end{equation}
can be introduced in terms of which the Lie bracket takes the form
of a direct sum \begin{equation}
\left[J_{a}^{\pm},J_{b}^{\pm}\right]=\epsilon_{abc}J_{\pm}^{c},\quad \left[J_{a}^{\pm},J_{b}^{\mp}\right]=0.\label{alg of so(2,2)}
\end{equation}
Thus, the Lie algebra in this case is $\ch\oplus \ch$. 
For later use we also note that  with $J_a=J_a^++J_a^-$  and   $\Pi_a=\sqrt{\lambda}J_a^+$   the brackets take
the form of a semidirect sum:
\begin{equation}
 \left[J_{a},J_{b}\right]=\epsilon_{abc}J^{c},\;
 [J_a,\Pi_b]=\epsilon_{abc}\Pi^c \mbox{ and } [\Pi_a,\Pi_b]=\sqrt{\lambda}\epsilon_{abc}\Pi^c,
\label{com rel for Pi} \end{equation}
which contracts to the Euclidean or Poincar\'e Lie algebra as $\lambda \rightarrow 0$.

\subsection{Iwasawa decomposition}
In all cases  except for the Euclidean situation with $\lambda >0$ (and hence $\Lambda_C>0$), we can introduce a vector  $ \mathbf{n}=(n^{0},n^{1},n^{2})\in \R^3 $ satisfying \begin{equation}
\mathbf{n}^{2}=\eta_{ab}n^{a}n^{b}=-\lambda, \label{the vector n} \end{equation}
and define the generator $\PT_a$ by  
\begin{equation}
 \PT_a=P_a+\epsilon_{abc}n^bJ^c.
\label{Iwa decomp}
\end{equation}
Then the Lie brackets on $\cg_\lambda$ take the form 
\begin{equation}
\left[J_{a},J_{b}\right]=\epsilon_{abc}J^{c},\;\left[J_{a},\PT_{b}\right]=\epsilon_{abc}\PT^{c}+n_bJ_a-\eta_{ab}(n^cJ_c),\;\;\left[\PT_{a},\PT_{b}\right]=n_a\PT_b-n_b\PT_a.\label{Iwa brackets}
\end{equation}
This shows in particular that both $\ch$ and the span of $\{\PT_{0},\PT_{1},\PT_{2}\}$
form Lie subalgebras of $\cg_{\lambda}.$ The latter depends on the
choice of the vector $\bv n.$ For $\mathbf{n}=0,$ which requires $\lambda=0,$
it is simply $\R^{3}$ with the trivial Lie bracket. For $\mathbf{n}\neq0,$
we will denote this Lie algebra by $\mathfrak{an}(2)_{\mathbf{n}}$ or, when the
dependence of $\bv n$ need not be emphasised, simply by $\mathfrak{an}(2).$ The decomposition
\begin{equation}
\cg_{\lambda}=\ch \oplus \mathfrak{an}(2)_{\bv n}, \end{equation}
implied by (\ref{Iwa decomp}), generalises the Iwasawa decomposition of $\mathfrak{so}(3,1)\simeq\mathfrak{sl}(2,\C)$
into a compact part $\mathfrak{su}(2)$ and  a non-compact part consisting of traceless, complex upper triangular matrices with real  diagonal. In that context, the corresponding Lie group  $AN(2)$ is the group of  $2\times 2 $ matrices of the form
\begin{equation}
\label{an2para}
\begin{pmatrix}
e^{\alpha} & \xi+i\eta \\
0 & e^{-\alpha} \\
 \end{pmatrix},\quad \alpha ,\xi,\eta \in \R,
 \end{equation}
 and the notation refers the abelian and the nilpotent parts of this group, 
which is isomorphic to the semidirect product $\R \rcross \R^2.$

\subsection{Quaternionic description} 
In \cite{Meusburger3}, the family of Lie algebra $\cg_\lambda$ is described in a unified fashion by identifying them with the 3d rotation and Lorentz Lie algebra over 
a commutative ring $R_\lambda.$  The ring $R_\lambda $ is a generalisation of the complex numbers, and   consists of elements of the form $a+\theta b,\; a,b,\in \R$ for a formal parameter $\theta$ satisfying $\theta^2=\lambda$. We call $a$ and $b$ the real and imaginary parts, and write
\begin{equation}
\text{Re}_\theta(a+\theta b)=a\quad \text{Im}_\theta(a+\theta b)=b \quad \forall a,b\in \R. 
\end{equation}
The addition in  $R_\lambda $  is the vector space addition in $\R^2$ and  multiplication rule is 
\begin{equation}
 (a+\theta b)\cdott (c + \theta d)=(ac+\lambda bd) + \theta (ad+bc)\quad \forall a,b,c,d\in \R.
\end{equation}
 There is a $\R-$linear involution $^*:R_\lambda \mapsto R_\lambda,$  
 defined via $(a+\theta b)^*=a-\theta b$,  called $\theta$-conjugation in the following. 
The ring $R_\lambda$ is a field in the case $\lambda < 0$ (the complex numbers) but has zero divisors when $\lambda \geq 0.$
 

As shown in \cite{MeusburgerSchroers6}, a convenient description of the local isometry groups  and their Lie algebras  in 3d gravity  can be  obtained in terms of unit (pseudo-) quaternions 
over the ring $R_\lambda$.  
We denote the unit imaginary (pseudo-) quaternions by   $e_a,\;a=0,1,2$. They satisfy the relations 
\begin{equation}
 e_ae_b=-\eta_{ab}+\epsilon_{abc}e^c, \quad a,b =0,1,2.
\end{equation}

 Quaternionic conjugation acts trivially on the identity $1$ (often omitted when writing quaternions) and acts on imaginary quaternions according to
\begin{equation}
\label{conjugate}
\bar{e}_a= -e_a,\qquad  a =0,1,2.
\end{equation}
It 
 is extended linearly to a general (pseudo-) quaternions,  which have the form 
\begin{equation}
v=v_3 + v^ae_a, \qquad v_0,v_1,v_2,v_3 \in \RR.
\end{equation}
 If we need to distinguish between the Euclidean and the Lorentzian situation  we write $\HH^E$ for 
 the set of quaternions (where $\eta_{ab}$ is the Euclidean metric) and $\HH^L$ for 
 the set of pseudo-quaternions (where $\eta_{ab}$ is the Minkowski metric), but we drop the superscript in expression which makes sense in either case.
The set of unit (pseudo) quaternions is  defined via
\begin{equation}
\HH_1=\{v\in \HH | v \bar{v} =1\}.
\end{equation}
It is easy to check that $\HH^E_1\cong SU(2)$ and $\HH_1^L\cong SL(2,\RR)$, which motivates our notation $\mathfrak{h}$ for the Lie algebra of either of these groups.

Combining the ring $R_\lambda$ with quaternions, we define
\begin{equation}
\mathbb{H}^{E,L}(R_{\lambda}):=\mathbb{H}^{E,L}\otimes_{\mathbb{R}}R_{\lambda},\end{equation}
whose elements  are of the form 
\begin{equation}
g=q_{3}+\theta k_{3}+(\mathbf{q}+\theta\mathbf{k})\cdott e,\quad q_{3},k_{3}\in\mathbb{R},\mathbf{q},\mathbf{k}\in\mathbb{R}^{3}. \end{equation}
As shown in \cite{MeusburgerSchroers6}, the local isometry groups in $3d$ gravity are isomorphic
to the multiplicative groups 
\begin{equation}
\mathbb{H}_{1}^{E,L}(R_{\lambda}):=\{g\in\mathbb{H}^{E,L}(R_{\lambda})|g\bar{g}=1\}\end{equation}
of unit (pseudo) quaternions over the commutative ring $R_{\lambda}$.  The  following isomorphisms hold:\begin{equation}\begin{array}{rl}
\mathbb{H}^{E}(R_{\lambda>0})&\cong SU(2)\times SU(2),\qquad\mathbb{H}_{1}^{L}(R_{\lambda>0})\cong SL(2,\mathbb{R})\times SL(2,\mathbb{R}),\\[1.0ex]
\mathbb{H}_{1}^{E}(R_{\lambda=0})&\cong   SU(2) \rcross \mathbb{R}^{3}\,\qquad\qquad\mathbb{H}_{1}^{L}(R_{\lambda=0})\cong  SL(2,\mathbb{R})  \rcross \mathbb{R}^{3}, \\[1.0ex]
\mathbb{H}_{1}^{E}(R_{\lambda<0})&\cong SL(2,\C),\qquad\quad \qquad\mathbb{H}_{1}^{L}(R_{\lambda<0})\cong SL(2,\C).
\end{array}\end{equation}
  The Lie algebra generators  $P_a$ and $J_a$  of $\mathfrak{g}_\lambda$ can  then be realised as 
\begin{equation}
 J_a=\frac{1}{2}e_a,\quad P_a=\theta J_a,
\end{equation}
 which reproduces the  brackets \eqref{eq:lie algebra}. 
In summary, we have the  following isomorphisms: 
\begin{equation}
\begin{array}{rl}
\cg_{\lambda}^{E}&\cong \left\{ \begin{array}{lll}
\mathfrak{su}(2)\oplus \mathfrak{su}(2) & \, \quad \quad \quad\mbox{for \ensuremath{\lambda>0}}\\
\mathfrak{iso}(3) &\, \quad  \quad \quad\mbox{for \ensuremath{\lambda=0}}\\
\mathfrak{so}\left(3,1\right) & \, \quad \quad \quad \mbox{for \ensuremath{\lambda<0}}\end{array}\right.\\[3.0ex]
\cg_{\lambda}^{L}&\cong \left\{ \begin{array}{lll}
\mathfrak{sl}\left(2,\mathbb{R}\right)\oplus \mathfrak{sl}\left(2,\mathbb{R}\right) & \quad\mbox{for \ensuremath{\lambda}\ensuremath{>0}}\\
\mathfrak{iso}(2,1) & \quad\mbox{for \ensuremath{\lambda=0}}\\
\mathfrak{so}\left(3,1\right) & \quad\mbox{for \ensuremath{\lambda<0}}\end{array}\right.\end{array}
\end{equation}

 \subsection{Parametrising unit (pseudo-)quaternions}


 For the calculations in this paper involving $\lambda \neq 0$, it is useful to introduce a basis of the unit quaternions which is adapted to the sign of $\lambda$. The construction we are about to give should be thought of as a generalisation of the parametrisation of $SU(2)$ and $SL(2,\RR)$ as subsets of $\R^4$ satisfying a constraint:
 \begin{equation}
\begin{pmatrix} a+i b & c+id \\-c+id & a-ib \end{pmatrix} \in SU(2) \quad \text{iff} \;\; a^2+b^2+c^2 +d^2=1,
 \end{equation}
 and 
 \begin{equation}
 \begin{pmatrix} a+b & c+d \\-c+d & a-b \end{pmatrix} \in SL(2,\R) \quad \text{iff}\; \; a^2-b^2+c^2 -d^2=1.
 \end{equation}
 For the required generalisation we need to distinguish cases according to the sign of the cosmological constant $\Lambda_C$. In all calculations  involving the vector $\bn$ and the condition \eqref{the vector n}  we use
$\mp$ or $\pm$ with  the upper sign referring to $\Lambda_C <0$, i.e. 
 $\lambda<0$  with Euclidean signature or $\lambda>0$  with Lorentzian signature, and  the lower sign referring to  $\Lambda_C>0$, i.e. $\lambda<0$ with Lorentzian signature (for Euclidean signature  and $\Lambda_C \geq 0$  there is no non-trivial solution of \eqref{the vector n}).  The Lorentzian  case with  $\lambda =0$ is considered separately.

For our construction we complement the vector   $\bv n\in \R^3$ appearing  in (\ref{the vector n}) by an orthogonal vector $\mathbf{m}$, satisfying $\mathbf{m}^{2}=\mp\frac{  1}{  \lambda}$ so that 
$|\mathbf{m}\times \mathbf{n}|^{2}=\pm1.$                                                                                                 
Then we introduce the quaternion  basis 
\begin{equation}
\label{adaptedbasis}
\left\{ 1,\mathbf{n\cdott e},(\mathbf{m\times n})\cdott\mathbf{e},\mathbf{m}\cdott\mathbf{e}\right\} 
\end{equation}
 and note that   any element $v\in\mathbb{H}_{1}$ can be written as 
\begin{equation}
v=a-b\mathbf{n}\cdott\mathbf{e}+c(\mathbf{m}\times\mathbf{n})\cdott\mathbf{e}+\lambda d\mathbf{m}\cdott\mathbf{e},\label{eq:v1}\end{equation}
where $a,b,c,d\in\mathbb{R}$, 
provided $v\bar{v}=1$, i.e.
\begin{equation}
a^{2}-\lambda b^{2}\pm(c^{2}-\lambda d^{2})=1.
\end{equation}
The elements of \eqref{adaptedbasis} satisfy the following algebraic relations
 \begin{equation}
\begin{array}{rl}
(\mathbf{n}\cdott\mathbf{e})^{2}&=\lambda, \\[1.0ex] 
((\mathbf{m}\times\mathbf{n})\cdott\mathbf{e})^{2}&=\mp1,\\[1.0ex] 
 (\bv m \cdott \bv e )^2&=\pm \frac{1}{\lambda},\\[1.0ex]
(\mathbf{n}\cdott\mathbf{e})(\mathbf{m}\cdott\mathbf{e})&=-(\mathbf{m}\cdott\mathbf{e})(\mathbf{n}\cdott\mathbf{e})=-(\mathbf{m\times n})\cdott\mathbf{e},\\[1.0ex]
(\mathbf{m}\cdott\mathbf{e})((\mathbf{m\times n})\cdott\mathbf{e})&=-((\mathbf{m\times n})\cdott\mathbf{e)(m\cdott e)}=\pm\frac{  1}{  \lambda}\mathbf{n}\cdott\mathbf{e},\\[1.0ex]
(\mathbf{n}\cdott\mathbf{e})((\mathbf{m\times n})\cdott\mathbf{e})&=-((\mathbf{m\times n}),\cdott\mathbf{e)(n\cdott e)} =-\lambda\mathbf{m}\cdott\mathbf{e}.\end{array}\label{eq:prop 1}\end{equation}
It follows that the non-trivial commutators between basis elements are 
\begin{equation}\begin{array}{rl}
 [\bv m \cdott \bv e,\bv n\cdott \bv e]&=2(\bv m \times \bv n)\cdott \bv e, \\
\left[\bv n \cdott \bv e,(\bv m \times \bv n) \cdott \bv e \right]&=-2 \lambda \bv m \cdott \bv e,\\
\left[\lambda \bv m \cdott \bv e,(\bv m \times \bv n)\cdott \bv e \right]& =\pm 2 \bv n \cdott \bv e.
\end{array}
\label{comrelmnmn}
\end{equation}

 When $\lambda=0$ and $\mathbf{n}^{2}=0$ with $\mathbf{n}\ne 0$
(i.e in the Lorentzian case), a second lightlike vector $\tilde{\mathbf{n}}$ is
introduced which satisfies
\begin{equation}
\tilde{\mathbf{n}}\cdott\tilde{\mathbf{n}}=0,\mbox{ }\mathbf{n}\cdott\mathbf{\tilde{n}}=1.\end{equation}
Then the vector $\tilde{\mathbf{m}}=\tilde{\mathbf{n}}\times\mathbf{n}$
is spacelike, with $\tilde{\mathbf{m}}\cdott\tilde{\mathbf{m}}=-1$.
These vectors are then used to parametrise elements $v\in\mathbb{H}_{1}$ via
\begin{equation}
v=a+b\,\mathbf{\tilde{m}}\cdott\mathbf{e}+\gamma\,\mathbf{n}\cdott\mathbf{e}+\tilde{\gamma}\,\tilde{\mathbf{n}}\cdott\mathbf{e}\end{equation}
in terms of $a,b,\gamma,\tilde{\gamma}\in\mathbb{R}$. The condition
$v\bar{v}=1$ is equivalent to\begin{equation}
a^{2}-b^{2}+2\gamma\tilde{\gamma}=1.\end{equation}
The  basis elements  in this case satisfy the relations
 \begin{equation}
\begin{array}{rl}
(\mathbf{\tilde{m}}\cdott\mathbf{e})^{2}&=1,\\
(\mathbf{\tilde{n}}\cdott\mathbf{e})(\mathbf{n}\cdott\mathbf{e})&=-1+\mathbf{\tilde{m}}\cdott\mathbf{e},\\
(\mathbf{n}\cdott\mathbf{e})(\mathbf{\tilde{n}}\cdott\mathbf{e})&=-1-\mathbf{\tilde{m}}\cdott\mathbf{e},\\
(\mathbf{\tilde{n}}\cdott\mathbf{e})(\mathbf{\tilde{m}}\cdott\mathbf{e})&=\mathbf{-(\tilde{m}}\cdott\mathbf{e)}(\mathbf{\tilde{n}}\cdott\mathbf{e})=\mathbf{\tilde{n}}\cdott\mathbf{e},\\
(\mathbf{\tilde{m}}\cdott\mathbf{e})(\mathbf{n}\cdott\mathbf{e})&=-(\mathbf{n}\cdott\mathbf{e})(\mathbf{\tilde{m}}\cdott\mathbf{e})=\mathbf{n}\cdott\mathbf{e}.\end{array}\label{eq:prop 2}\end{equation}
 with commutation relations 
\begin{equation}\begin{array}{rl}
  \left[\tilde{\bv n} \cdott \bv e,\bv n \cdott \bv e\right] &=2\tilde{\bv m}\cdott \bv e, \\
 \left[\tilde{\bv n} \cdott \bv e,\tilde{\bv m} \cdott \bv e\right]&=2 \tilde{ \bv n}\cdott \bv e, \\
\left[\tilde{\bv m} \cdott \bv e,\bv n \cdott \bv e\right]&=2 \bv n\cdott \bv e.
\end{array}\label{comrelntntm}
 \end{equation}

\subsection{Model spacetimes}
For the calculations in this paper we will also require some background on the model spacetimes of 3d gravity. As reviewed in the introduction, these depend on the cosmological constant and the signature. 
A unified description of the model spacetimes  in quaternionic  language was given in \cite{MeusburgerSchroers6}. Here we only note the conclusion that the model spacetimes can be realised as hypersurfaces in an ambient $\R^4$,  whose coordinates we denote $w_0,w_1,w_2,w_3$. 
Then, for given signature and $\lambda$, the model spacetimes are 
\begin{equation}
\label{modelspacetime}
 W_\lambda=\{(w_3, \bv w )\in \R^4|w_3^2+\lambda \bv w^2=1\},
 \end{equation}
 with a metric induced by the quadratic form defining the constraint 
 \begin{equation}
 \label{constraint}
 w_3^2+\lambda \bv w^2=1.
 \end{equation}
 As explained in \cite{MeusburgerSchroers6},  it is natural to think of the ambient $\R^4$ itself as the subspace of $\mathbb{H}(R_\lambda)$ which is fixed under simultaneous quaternionic and $\theta$-conjugation.  This can be made manifest by writing 
  \begin{equation}
 \label{wpara}
 w=w_0 +\theta\mathbf{w}\cdott\mathbf{e},
\end{equation}
which we will use later in this paper.

 In  the Euclidean case,  $W_\lambda$ is  two copies of hyperbolic space when $\lambda<0$, two copies of Euclidean space (embedded as  affine spaces in $\R^4$) when $\lambda=0$ and the 3-sphere when $\lambda >0$.  In  the Lorentzian case,   $W_\lambda$ is a double cover of de Sitter space space when $\lambda<0$, two copies of Minkowski  space (embedded as  affine spaces in $\R^4$) when $\lambda=0$ and a double cover of anti-de Sitter space when $\lambda >0$.
 
\subsection{Parametrisations of the $AN(2)$ group}
Finally, we review two parametrisations of elements in the subgroup $AN(2)_\bn$ of $\mathbb{H}_1(R_\lambda)$ which is obtained by exponentiating  the Lie algebra $\mathfrak{an}(2)_\bn$. Both parametrisations use  quaternionic  notation, and we refer to  \cite{MeusburgerSchroers6} for details and motivation. We omit the subscript $\bn$  when  the dependence on $\bn$  is not important. 

In the 
first parametrisation, an element $r\in AN(2)$ is given in terms of
an unconstrained vector $\mathbf{q}\in\mathbb{R}^{3}$ by
\begin{equation}
r=\sqrt{1+(\bv q\cdott \bv n)^2/4}+ \frac{\theta}{2}\bv q \cdott \bv e+\frac{1}{2}\bv q \times \bv n\cdott \bv e.
\end{equation}
 In the second, an element $r\in AN(2)$ takes the form\begin{equation}
r(\alpha,z)=(1+zQ)e^{\alpha N},\end{equation}
where $z\in R_{\lambda}$ and $\alpha\in\mathbb{R}$, and $N$ and $Q$ are defined below.  This should be thought of as a generalisation of the matrix parametrisation \eqref{an2para}. 

For the case $\lambda\neq0,$
\begin{equation}
N=-\frac{  1}{  \theta}\mathbf{n}\cdott\mathbf{e}, \quad  Q=\frac{  \theta}{  2}\mathbf{m}\cdott\mathbf{e}+\frac{  1}{  2}(\mathbf{m}\times\mathbf{n})\cdott\mathbf{e}.
\end{equation}
Thus on setting $z=\xi+\theta\eta,$ with $\xi,\eta\in\R,$ we get
\begin{align}
r(\alpha,z)&=\cosh\alpha+e^{-\alpha}\left(\frac{  \lambda\eta}{  2}\mathbf{m}\cdott\mathbf{e}+\frac{  \xi}{  2}(\mathbf{m}\times\mathbf{n})\cdott\mathbf{e}\right)\nonumber \\[1.0ex]
&+\theta\left(-\frac{  1}{  \lambda}\mathbf{n}\cdott\mathbf{e}\sinh\alpha+e^{-\alpha}(\frac{  \xi}{  2}\mathbf{m}\cdott\mathbf{e}+\frac{  \eta}{  2}(\mathbf{m}\times\mathbf{n})\cdott\mathbf{e})\right).
\label{eq:r left action}
\end{align}
The parameters $\mathbf{q}$ and $(\alpha,\xi,\eta)$ are related by \begin{equation}
\mathbf{q}=-\frac{2}{\lambda}\sinh\alpha\mathbf{n}+e^{-\alpha}\xi\mathbf{m}+e^{-\alpha}\eta(\mathbf{m}\times\mathbf{n}).\label{eq:q}\end{equation} 

For $\lambda=0$, an element $r\in AN(2)$ is parametrised
by $\mathbf{q}\in\mathbb{R}^{3}$ such that\begin{equation}
\mathbf{q}=(e^{2\alpha}-1)\tilde{\mathbf{n}}+e^{-\alpha}\xi\tilde{\mathbf{m}}+e^{-\alpha}\eta\mathbf{n} \label{eq: q0} \end{equation}
and \begin{equation}
r(\alpha,z)=\cosh\alpha+\mathbf{m}\cdott\mathbf{e}\sinh\alpha+e^{-\alpha}\frac{\xi}{2}\mathbf{n}\cdott\mathbf{e}+\theta\left(\tilde{\mathbf{n}}\cdott\mathbf{e}\sinh\alpha+e^{-\alpha}(\frac{\xi}{2}\mathbf{\tilde{m}}\cdott\mathbf{e}+\frac{\eta}{2}\mathbf{n}\cdott\mathbf{e})\right).\end{equation}
Finally,  we note that the coordinates $\alpha,\xi,\eta$  may also be viewed as differentiable  functions 
\begin{equation}
\label{Fun AN}
f: AN(2)\mapsto \R.                                                                     
\end{equation}
This is the viewpoint which we mostly adopt in this paper.

\section{Bicrossproducts and semidualisation}
\subsection{General formalism }
\label{method}
The calculations in this paper are based on a  general  method for constructing  bicrossproduct Hopf algebras from  factorisable Lie groups. The construction is a particular instance of the procedure of semidualisation  of Hopf algebras.  In the case at hand, the initial Hopf algebra is the universal enveloping algebra of one of  the Lie algebras arising in 3d gravity. We briefly review those features  of the construction which are required in the application to  the local isometry Lie algebras of 3d gravity, and refer the reader to  the original literature \cite{MajidPhD,Majidbicross} or the book \cite{Majidbook1} for a wider discussion of semidualisation. General background on Hopf algebras can be found in \cite{Majidbook1} or \cite{CP}.

Suppose that $X$ is  a group (not necessarily a Lie group) which can be factorised into subgroups $G,M\subset X$ such that $X=GM $
and  that the map $G\times M\rightarrow X$ given by multiplying within $X$ is
a bijection. Then   every element of $x\in X$ can be uniquely
expressed as a normal ordered product $x=gm$ of the elements $g\in G, m \in M.$ The unique factorisation allows one 
to define  a left action $\la$ of $M$ on $G$ and a right action $\ra$ of $G$ on $M$
\begin{equation}
\begin{array}{cc}
\la:M\times G\rightarrow G, & \ra:M\times G\rightarrow M\end{array}
\end{equation}
by starting with elements $g\in G,m\in M, $ multiplying them in `wrong order' and then factorising: 
\begin{equation} 
mg=(m\la g)(m\ra g), \quad \forall g\in G,m\in M.\end{equation}
These actions obey
\begin{eqnarray}
e\la g&=g, \quad (m_{1}m_{2})\la g &=m_{1}\la(m_{2}\la g),\\
m\la e&=e, \quad  m\la (g_{1}g_{2}) &=(m\la g_{1})((m\ra g_1)\la g_2),\\
e\ra g&=e, \quad  m \ra (g_{1}g_{2})&=(m\ra g_1)\ra g_2,\\
m\ra e&=m,  \quad (m_{1}m_{2})\ra g &= (m_1\ra (m_2\la g ))(m_2 \ra g),
\end{eqnarray}
for $m_{1},m_{2}\in M$, $g_{1},g_{2}\in G$, with  $e$ denoting
the relevant group identity element.  If we want to emphasise the actions of the subgroups $G$ and $M$ of $X$ on each other, we say that $X$ is the  \emph{double cross product group} $X=G\bowtie M$.

Generally,  groups $G$ and $M$  with actions on each other with the above properties  are called a matched pair. Given a matched pair one can define the double cross product
 $G\bowtie M$  as the set $G\times M$ with product 
 \begin{equation}
(g_{1},m_{1})\cdott(g_{2},m_{2})=\left(g_{1}(m_{1} \la g_{2}),(m_1\ra g_{2}) m_{2}\right),\label{eq:double cross product}
\end{equation}
unit $e=(e,e)$ and inverse
 \begin{equation}
 (g,m)^{-1}=(m^{-1}\la g^{-1},m^{-1}\ra g^{-1})
 \end{equation}
 and  with $G,M$ as subgroups.

As an aside we note that the above construction can be extended into the quantum group setting. 
Suppose then that $(H_{1},H_{2})$ are a matched pair of quantum groups with $H_{1}\bowtie H_{2}$ the associated double
cross product. Then there is another quantum group denoted by $H_{2} \rlbicross H_{1}^{\ast},$ where $H_{1}^{\ast}$ is the dual of $H_1,$
called semidualisation of the matched pair.   Its dual $H_{2}^\ast \lrbicross H_{1}$ is another semidualisation, and the one we will use in this paper.  Again we refer to \cite{Majidbook1} for  a detailed construction.

Consider now  a matched pair of groups,  $M$ and $G$. Assuming initially that  both $M$ and $G$  are finite groups, we  write  $\C G$  for the  group algebra of $G$ and $\C(M)$ for  the   space   of functions on $M$. The construction of the bicrossproduct Hopf algebra $\C(M) \lrbicross \C G$   then proceeds as follows.
The vector space underlying the bicrossproduct $\C(M) \lrbicross \C G$ is the  
tensor product $\C(M) \otimes \C G $.  Let $g\in G$ and $f\in \C(M)$ and consider 
elements of the form $f\otimes g\in \C(M)\otimes \C G$. The algebra has the multiplication
$\bullet$, which is ordinary multiplication of the group elements
and pointwise multiplication of the functions, twisted by the right
action $\ra$:
\begin{equation}
f_{1}\otimes g_{1}\bullet f_{2}\otimes g_{2}\left(m\right)= f_{1}\left(m\right)f_2\left(m\ra g_{1}\right)
\otimes g_{1}g_{2}.
\label{bi product 1}
\end{equation}
The unit  is $e\otimes1,$ where $1$ is the function which
is $1$ everywhere on $M.$ 

In order to characterise the co-algebra structure we need to give 
the co-product and the co-unit. The co-product is the usual co-multiplication
for the functions on the group $M$,  but the co-product of a group
element $g\in G$ is twisted by the left action $\la$:
\begin{equation}
\Delta\left(f\otimes g\right)\left(m_{1},m_{2}\right)=f\left(m_{1}m_{2}\right)
\otimes (m_{2}\la g)\otimes g.\label{bi coproduct 1}\end{equation}
The co-unit is $\epsilon(f\otimes g)=f(e)$. Note that we may identify
$(\C(M) \otimes \C G)\otimes(\C(M) \otimes \C G)$ with
$ \C(M\times M)\;\otimes \C\;\; G\times G$ as  a vector space and this
is what has been done on the right hand side of (\ref{bi coproduct 1})   for ease of notation.
The antipode is  \begin{equation}
S(f\otimes g)(m)= f\left(m^{-1}\ra g^{-1}\right)\otimes
\left( m^{-1} \la g\right)^{-1}.
\label{bi antipode 1}
\end{equation}

 The bicrossproduct   construction can be generalised  to Lie group setting \cite{MajidPhD,MajidNonComST,  Majid90} under the weaker (and more useful) assumption that 
the original group  $X$ is factorisable into the subgroups  $G$ and $ M$  near the identity. The precise formulation requires that  one defines the group algebra of a Lie group, for which there are various options, satisfying different analytical requirements.  We will by-pass these issues here since we are only interested in the infinitesimal construction. More precisely, starting with a  double cross sum $\cg\bowtie\cm$ of Lie algebras, we construct the semidual  of the universal enveloping algebra  
$U(\cg \bowtie \cm) = U(\cg) \bowtie U(\cm)$ as the  quantum group $ \C (M)\lrbicross U(\cg) $, where $\C(M)$ are complex-valued smooth functions on $M$,  dual to $U(\cm).$  In practice, this means applying the above  expressions for the  Hopf algebra structures   $\C(M) \lrbicross \C G $ to  Lie groups $G$ and $M$, with 
\begin{equation}
g=e^{\varepsilon \chi}, \quad \chi\in\cg, 
\end{equation}
expanding
\begin{equation}
   g=1+\varepsilon \chi+O(\varepsilon^2),\label{exp of g}
\end{equation}
and keeping only leading, linear terms in $\varepsilon$.

Using (\ref{exp of g}), we obtain from (\ref{bi product 1})
\begin{align}
[f \tens  1 , 1 \tens \chi ](m) &=\left.\frac{ d }{ d\varepsilon}\right|_{\varepsilon=0}
\left(  \left(f(m)-f( m\ra e^{\varepsilon \chi}\right)  \otimes  e^{\varepsilon\chi}  \right), \nonumber \\
&= - \left.\frac{ d }{ d\varepsilon}\right|_{\varepsilon=0}f( m\ra e^{\varepsilon \chi}) \otimes 1,\nonumber \\
\left[ f_1\tens 1 ,f_2\tens 1 \right] & = 0,
\label{bi productinfini}
\end{align}
together with the commutation relations of the Lie algebra generators $\chi \in \cg$.  The co-products (\ref{bi coproduct 1}) become
\begin{equation}
\Delta (f\tens \chi)(m_1,m_2) = f(m_1 m_2) \tens
\left.\frac{ d }{ d\varepsilon}\right|_{\varepsilon=0}
\left( m_2\la e^{\varepsilon \chi}\right)\tens \chi.
\label{bi coproduct}
\end{equation}
Finally, the antipode is  
\begin{equation}
S\left(f\otimes \chi\right)(m)=\left.
\frac{ d }{ d\varepsilon}\right|_{\varepsilon=0}
\left(  f\left(m^{-1}\ra e^{-\varepsilon\chi}\right) \otimes \left( m^{-1} \la e^{\varepsilon\chi}\right)^{-1} \right).
\label{bi antipode}
\end{equation}

\subsection{Simple examples}

\label{simpleexamples}
We  consider some special cases of the general construction above, which we later apply to some of the local isometry groups in $3d$ gravity. 
Let $G$ be a finite group and take $X=G\times G$, but viewed as a semidirect product $G\rcross G$.
 This is the group analogue of the Lie algebra decomposition \eqref{com rel for Pi} of a direct sum; 
 see \cite{MajidSchroers} for details. 
 Then the bicrossproduct construction of the previous subsection, with $M=G$ and  right  action
 \begin{equation} h \ra  g  = g^{-1}h g
 \end{equation}
 as well as the  trivial left action $h\la g = g$ for $h,g\in G$ gives the quantum double 
  $D(G)$. It has the algebra structure $\C (G) \lcross \C G$ and the  direct co-product
  \begin{equation}
  \Delta f\otimes g \,(h_1,h_2) =f(h_1h_2) \;g\otimes g.
  \end{equation}
  This is the simplest example of an interesting non-commutative and non-cocommutative Hopf algebra in the case where $G$ is non-commutative.
If $G$ is a Lie group with Lie algebra $\mathfrak{g}$, then one can define  the quantum double $D(U(\cg))=\C (G) \lcross  U(\cg) $
 by differentiating near the identity element in $G$, as illustrated above for general bicrossproducts.  

Another example is the semidirect product $X= GT$ where $G$ and $T$ are Lie subgroups of the Lie group $X$ and  $T$ is abelian.   In this case, semidualisation leads to the Hopf algebra $T^* \lcross U(\cg) $, where $T^*$  is the Pontryagin dual group (group  of characters) of $T$, see
 \cite{Majidbook1}.

\section{Double cross product structure  of local isometry groups }

\subsection{Factorisations  of local isometry groups}
Table~\ref{doublecrosstable} provides a list of the local isometry groups arising in 3d gravity with their corresponding 
matched pairs of right cross products  or double cross products. 
\begin{table}[H]
\centering
\begin{tabular}{|l|c|c|}
\hline
&&\\
& Euclidean signature   & Lorentzian signature\\
&&\\
\hline
&&\\
$ \lambda>0$ & $\widetilde{SO}(4)= SU(2) \rcross SU(2) $   & $\widetilde{SO}(2,2)=\left\{ \begin{array}{ll} SL(2,\R)\rcross SL(2,\R) \\ SL(2,\R)\dcross_s AN(2) \end{array}\right. $\\ 
&&\\
 \hline
&&\\
$ \lambda=0$ & $\tilde{E}_3=SU(2) \rcross  \R^3  $ & $\tilde{P}_3=\left\{ \begin{array}{ll} SL(2,\R )  \rcross \R^3    \\ SL(2,\R)\dcross_l AN(2)  \end{array}\right. $    \\
&&\\
\hline
&&\\
$ \lambda<0$ & $ SL(2,\C)=SU(2) \dcross AN(2) $& $SL(2,\C)=SL(2,\R)\dcross_t AN(2)$\\
 
&&\\

\hline
\end{tabular}
 \vspace{.5cm}
\caption{Local isometry groups  in 3d gravity and their factorisations}
\label{doublecrosstable}
 \end{table}
Starting with Euclidean signature,  the local isometry group is the `prototype' double cross product $SL(2,\C)=SU(2)\dcross AN(2)$, analysed in detail in \cite{Majidbook1}, in the case $\lambda<0$. 
 When $\lambda>0,$ the local isometry group $\widetilde{SO}(4)$  can also be viewed as  the semidirect product $\widetilde{SO}(4)= SU(2) \rcross SU(2)$ \cite{MajidSchroers}.
The corresponding Lie algebra $\mathfrak{su}(2) \rcross\mathfrak{su}(2) $ has generators $J_a$ and $\Pi_a,$ with commutation relation (\ref{com rel for Pi}).
When $\lambda=0,$ the local isometry group is the double cover of the Euclidean group with the canonical factorisation given in the table.

For Lorentzian signature, a family of  double cross product 
factorisations is implemented by (\ref{Iwa decomp}) and depends on the vector $\bv n,$ 
which  may be spacelike, lightlike or timelike. In Table~\ref{doublecrosstable} we use  
$\dcross_s, $ $\dcross_l$  and $\dcross_t $ to denote the double cross products with, respectively, a spacelike, lightlike
and timelike deformation vector  $\bn$. When  $\lambda>0$, the local isometry group $\widetilde{SO}(2,2)=SL(2,\R)\times SL(2,\R)$ can also be viewed as  the semidirect product $SL(2,\R) \rcross SL(2,\R)$, with 
 Lie algebra $\mathfrak{sl}(2,\RR) \rcross\mathfrak{sl}(2,\R)$ and generators $J_a$ and $\Pi_a,$ and commutation relation (\ref{com rel for Pi}). Finally,  we have, as a degenerate case when  $\lambda =0$ and $\bv n=0$,  the semidirect product $SL(2,\R)  \rcross  \R^3 $ with Poincar\'e Lie algebra given by (\ref{eq:lie algebra}).

\subsection{Right action of $U(\ch)$ on $\C(AN(2))$}

In the following, we compute the infinitesimal  right action $\ra $ of $U(\ch)$ on $AN(2)$  from the finite right action of $\mathbb{H}_1$ on $AN(2)$ given in \cite{MeusburgerSchroers6}, using the method outlined in Sect.~\ref{method}. We then  extract how the
functions on $AN(2)$ defined in (\ref{Fun AN}) transform under this action. 

Geometrically, the right action of $\mathbb{H}_1$ on $AN(2)$ is the pull-back of the natural action of 
$\mathbb{H}_1$ on the model spacetimes via a  map that identifies $AN(2)$ with (parts of ) the model spacetime. 
In terms of the parametrisation $\mathbf{q}\in\mathbb{R}^{3}$,  this identification is a map 
 \begin{equation}
 \label{Sdef}
 S:\mathbb{R}^3\rightarrow W_{\lambda},
 \end{equation}
where $W_\lambda$ is defined in \eqref{modelspacetime}, 
 with the explicit form
 \begin{equation}
S(\mathbf{q})=(1-\frac{\lambda}{2}\mathbf{q}^{2},\sqrt{1+\frac{(\mathbf{q}\cdott\mathbf{n})^{2}}{4}}\mathbf{q}+\frac{1}{2}\mathbf{q}\times(\mathbf{q}\times\mathbf{n})).
\label{eq:S}
\end{equation}
The conjugation action of elements $v$ of $\mathbb{H}_1$  on the model spacetime  is
  \begin{equation}
I_{v}:W_\lambda \rightarrow W_\lambda, \qquad I_{v}(w)=vw\bar{v},
\end{equation}
where we used the parametrisation \eqref{wpara} of elements $w\in W_\lambda$. 
It was shown in \cite{MeusburgerSchroers6}  that the inverse of $S$ generally only exists in a subset of $W_\lambda$, but the details will not concern us here. When it exists, the formula for the inverse is 
\begin{equation}
S^{-1}(w)=\frac{1}{\sqrt{w_{3}+\mathbf{w}\cdott\mathbf{n}}}\left(\mathbf{w}+\frac{\mathbf{w}^{2}}{1+w_{3}}\mathbf{n}\right).
\label{eq:inverse}
\end{equation}
Combining these maps, we obtain the promised right action of $\mathbb{H}_1$ on $AN(2)$:
\begin{equation}
\mathbf{q}\ra v=S^{-1}(I_{v^{-1}}(S(\mathbf{q})).
\end{equation}

When $\lambda\neq 0,$ substituting (\ref{eq:q}) into (\ref{eq:S}) gives \begin{equation}
S(\mathbf{q})=(w_{3},\xi\mathbf{m}+\eta(\mathbf{m}\times\mathbf{n})+\frac{1}{\lambda}(w_{3}-e^{2\alpha})\mathbf{n})\label{map S},\end{equation}
where\begin{equation}
w_{3}=\cosh2\alpha\pm\frac{1}{2}(\xi^{2}-\lambda\eta^{2})e^{-2\alpha}.\label{eq:w3}\end{equation}

In order to compute the commutators \eqref{bi productinfini} it is  sufficient to know the  infinitesimal right action of elements  $v\in\mathbb{H}_1$ near the identity.  We therefore  compute to linear  order  in the remainder of this section. We
write $v=1+\varepsilon\mathbf{n}\cdott\mathbf{e}$ and compute:
\begin{equation}
I_{v^{-1}}(w)=w_{3}+\theta(\mathbf{w}+2\varepsilon(\mathbf{w}\times\mathbf{n}))\cdott\mathbf{e}
+ O(\varepsilon^2),
\end{equation}
so that 
\begin{equation}
I_{v^{-1}}(S(\mathbf{q}))=w_{3}+\theta\left((\xi+2\varepsilon\lambda\eta)\mathbf{m}+(\eta+2\varepsilon\xi)(\mathbf{m}\times\mathbf{n})+\frac{1}{\lambda}(w_{3}-e^{2\alpha})\mathbf{n}\right)\cdott\mathbf{e} + O(\varepsilon^2),
\end{equation}
where $w_{3}$ is defined in (\ref{eq:w3}). The right action
$r\ra v$ for this case is  therefore
\begin{equation}
S^{-1}\left(I_{v^{-1}}(S(\mathbf{q}))\right)=e^{-\alpha}(\xi+2\varepsilon\lambda\eta)\mathbf{m}+e^{-\alpha}(\eta+2\varepsilon\xi)(\mathbf{m}\times\mathbf{n})-\frac{2}{\lambda}\sinh\alpha\mathbf{n}
+ O(\varepsilon^2).
\label{eq:r1}
\end{equation}
This gives a new element $r\in AN(2)$ in which, to leading order in $\varepsilon,$ \begin{equation}
\begin{array}{rl}
\alpha &\rightarrow\alpha,\\
\xi &\rightarrow\xi+2\varepsilon\lambda\eta,\\
\eta &\rightarrow\eta+2\varepsilon\xi.\end{array}\end{equation}

Similarly, 
for $v=1+\varepsilon(\mathbf{m}\times\mathbf{n})\cdott\mathbf{e}$, we find
\begin{equation}
I_{v^{-1}}(w)=w_{3}+\theta(\mathbf{w}+2\varepsilon(\mathbf{w}\times(\mathbf{m}\times\mathbf{n})))\cdott\mathbf{e} + O(\varepsilon^2),
\end{equation}
which gives 
\begin{equation}
I_{v^{-1}}(S(\mathbf{q}))=w_{3}+\theta\left((\xi-2\varepsilon(w_{3}-e^{2\alpha}))\mathbf{m}+\eta(\mathbf{m}\times\mathbf{n})+\frac{1}{\lambda}(w_{3}-e^{2\alpha}\pm2\varepsilon\xi)\mathbf{n}\right)\cdott\mathbf{e} + O(\varepsilon^2).
\end{equation}
The  infinitesimal  right action is therefore 
\begin{equation}
\begin{array}{rl}
r\ra v &=e^{-\alpha}\left(\xi+\varepsilon(e^{2\alpha}-e^{-2\alpha})\pm\varepsilon\lambda\eta^{2}e^{-2\alpha}\right)\mathbf{m}+e^{-\alpha}(\eta\pm\varepsilon\xi\eta e^{-2\alpha})(\mathbf{m}\times\mathbf{n})\\[1.0ex]
&-\frac{  2}{  \lambda}\left((1\pm\varepsilon\xi e^{-2\alpha})\sinh\alpha\mp\varepsilon\xi e^{-\alpha}\right)\mathbf{n} + O(\varepsilon^2).\end{array}\label{eq:r2}\end{equation}
Thus in this case, $r\ra v$ transforms the parameters of $r$ in the following manner, to linear order in $\varepsilon$:\begin{equation}
\begin{array}{rl}
\alpha &\rightarrow\alpha\mp\varepsilon\xi e^{-2\alpha},\\
\xi &\rightarrow\xi+\varepsilon e^{-2\alpha}\left((e^{4\alpha}-1)\mp(\xi^{2}-\lambda\eta^{2})\right)\\
\eta &\rightarrow\eta.\end{array},\end{equation}

Finally, for $v=1+\lambda\varepsilon\mathbf{m}\cdott\mathbf{e}$, we have
\begin{equation}
I_{v^{-1}}(w)=w_{3}+\theta(\mathbf{w}+2\varepsilon\lambda(\mathbf{w}\times\mathbf{m}))\cdott\mathbf{e} + O(\varepsilon^2),
\end{equation}
so that 
\begin{equation}
I_{v^{-1}}(S(\mathbf{q}))=w_{3}+\theta\left(\xi\mathbf{m}+(\eta-2\varepsilon(w_{3}-e^{2\alpha})(\mathbf{m}\times\mathbf{n})+\frac{1}{\lambda}(w_{3}-e^{2\alpha}\mp2\varepsilon\lambda\eta)\mathbf{n}\right)\cdott\mathbf{e} + O(\varepsilon^2), 
\end{equation}
and
\begin{equation}
\begin{array}{rl}
r\ra v&=e^{-\alpha}\left(\xi\mp\varepsilon\lambda\xi\eta e^{-2\alpha}\right)\mathbf{m}+e^{-\alpha}\left(\eta+\varepsilon(e^{2\alpha}-e^{-2\alpha})\mp\varepsilon\xi^{2}e^{-2\alpha}\right)(\mathbf{m}\times\mathbf{n})\\[1.0ex]
&-\frac{  2}{  \lambda}\left((1\mp\varepsilon\lambda\eta e^{-2\alpha})\sinh\alpha\pm\varepsilon\lambda\eta e^{-\alpha}\right)\mathbf{n}+ O(\varepsilon^2).\end{array}\label{eq:r3}\end{equation}
Therefore, $r\ra v$ in this case has the following parameters, to linear order in $\varepsilon$:
\begin{equation}
\begin{array}{rl}
\alpha &\rightarrow \alpha\pm\varepsilon\lambda\eta e^{-2\alpha},\\
\xi &\rightarrow\xi,\\
\eta &\rightarrow\eta+\varepsilon e^{-2\alpha}\left((e^{4\alpha}-1)\mp(\xi^{2}-\lambda\eta^{2})\right).\end{array}\end{equation}

Turning to the case  $\lambda=0,$ we obtain from (\ref{eq:S}) and (\ref{eq: q0}) 
\begin{equation}
S(\mathbf{q})=\left(1,(e^{2\alpha}-1)\tilde{\mathbf{n}}+\xi\tilde{\mathbf{m}}+e^{-2\alpha}(\eta+\frac{1}{2}\xi^{2})\mathbf{n}\right).\end{equation}
Suppose  $v=1+\varepsilon\mathbf{n}\cdott\mathbf{e}$, then
\begin{equation}
I_{v^{-1}}\left(S(\mathbf{q})\right)=1+\theta\left((e^{2\alpha}-1)\tilde{\mathbf{n}}+(\xi+2\varepsilon(e^{2\alpha}-1))\tilde{\mathbf{m}}+\left(e^{-2\alpha}(\eta+\frac{1}{2}\xi^{2})+2\varepsilon\xi\right)\mathbf{n}\right)\cdott\mathbf{e}.
\end{equation}
Therefore in $r\ra v$, the parameters $\alpha,\xi,\eta$  transform  infinitesimally according to  
\begin{equation}
\begin{array}{rl}
\alpha &\rightarrow\alpha,\\
\xi &\rightarrow\xi+2\varepsilon(e^{2\alpha}-1),\\
\eta &\rightarrow\eta+2\varepsilon\xi.\end{array}\end{equation}

For $v=1+\varepsilon\tilde{\mathbf{m}}\cdott\mathbf{e}$,
\begin{equation}
I_{v^{-1}}\left(S(\mathbf{q})\right)=1+\theta\left((1+2\varepsilon)(e^{2\alpha}-1)\tilde{\mathbf{n}}+\xi\tilde{\mathbf{m}}+e^{-2\alpha}\left((\eta+\frac{1}{2}\xi^{2})-2\varepsilon(\eta+\frac{1}{2}\xi^{2})\right)\mathbf{n}\right)\cdott\mathbf{e} + O(\varepsilon^2),
\end{equation}
transforming the parameters of $r$  infinitesimally as follows:
\begin{equation}
\begin{array}{rl}
\alpha &\rightarrow\alpha+\varepsilon(1-e^{-2\alpha}),\\
\xi &\rightarrow\xi,\\
\eta &\rightarrow\eta-2\varepsilon(\eta+\frac{  1}{  2}\xi^{2})e^{-2\alpha}.\end{array}\end{equation}
Finally, for $v=1+\varepsilon\tilde{\mathbf{n}}\cdott\mathbf{e}$
we have \begin{equation}
I_{v^{-1}}\left(S(\mathbf{q})\right)=1+\theta\left((e^{2\alpha}-1-2\varepsilon\xi)\tilde{\mathbf{n}}+\left(\xi-2\varepsilon e^{-2\alpha}(\eta+\frac{1}{2}\xi^{2})\right)\tilde{\mathbf{m}}+e^{-2\alpha}(\eta+\frac{1}{2}\xi^{2})\mathbf{n}\right)\cdott\mathbf{e} + O(\varepsilon^2)
\end{equation}
and the parameters of $r\in AN(2)$ under the right action transform infinitesimally according to
\begin{equation}
\begin{array}{rl}
\alpha &\rightarrow\alpha-\varepsilon\xi e^{-2\alpha},\\
\xi &\rightarrow\xi-2\varepsilon e^{-2\alpha}(\eta+\frac{  1}{  2}\xi^{2}),\\
\eta &\rightarrow\eta.\end{array}\end{equation}

\subsection{Left action of  $AN(2)$ on $U(\ch)$}
If $\lambda\neq0,$  we define the projections
 \begin{equation}
P=\frac{1}{2}(1+N)\mbox{ and }\bar{P}=\frac{1}{2}(1-N),\label{eq:ps}\end{equation}
where $N=-\frac{  1}{  \theta}\mathbf{n}\cdott\mathbf{e}.$ Then the left action of an element $r\in AN(2)$ on an element 
$v\in U(\ch)$ is given in \cite{MeusburgerSchroers6} 
as \begin{equation}
r\la v=\frac{1}{N\_}(rvP+r^{\ast}v\bar{P})\label{eq:l action}\end{equation}
 where the normalisation factor 
\begin{equation}
N_{\_}=| rvP+r^{\ast}v\bar{P}|\label{N}  \end{equation}
ensures that $r\la v$ is  a unit (pseudo) quaternion. We have written $|q|=\sqrt{q \bar q}$
for the `norm' of a quaternion here. In the Lorentzian case this norm could be ill-defined or zero, but in our applications  this will not concern us since  $v$ will be near the identity  (and hence has well-defined `norm') and we only consider infinitesimal changes.

Again we derive the infinitesimal version. Suppose $v=1+\varepsilon\mathbf{n}\cdott\mathbf{e}$, with $\varepsilon$ infinitesimal.  Then
putting (\ref{eq:r left action})  and (\ref{eq:ps})
into (\ref{eq:l action}) and using the properties in (\ref{eq:prop 1}),
we see that the  left action of $r$ on $v$ leaves $v$ invariant,
i.e. \begin{equation}
r\la v=1+\varepsilon\mathbf{n}\cdott\mathbf{e}.\label{eq:L act I}\end{equation}

Next, we take $v=1+\varepsilon(\mathbf{m}\times\mathbf{n})\cdott\mathbf{e}$.
The  left action (\ref{eq:l action}) then simplifies to \begin{equation}
r\la v=1+\varepsilon(\mathbf{m}\times\mathbf{n})\cdott\mathbf{e}\, e^{-2\alpha}\pm\varepsilon\mathbf{n}\cdott\mathbf{e}\, \eta \, e^{-2\alpha} + O(\varepsilon^2)
\label{eq:L act II}
\end{equation}
on using properties (\ref{eq:prop 1}). 
 Considering $v=1+\lambda\varepsilon\mathbf{m}\cdott\mathbf{e}$ the  left action of $r$ on $v$ can easily be obtained in
a similar fashion. Here, \begin{equation}
r\la v=1+\varepsilon\lambda\mathbf{m}\cdott\mathbf{e}\, e^{-2\alpha}\pm\varepsilon\mathbf{n}\cdott\mathbf{e}\, \xi \, e^{-2\alpha} + O(\varepsilon^2).
\end{equation}
 
Next we  consider $\lambda=0$ and $\mathbf{n}\neq 0$.
The  left action $r\la v$ of $r\in AN(2)$ on $v\in U(\ch)$ is
given in \cite{MeusburgerSchroers6} as 
\begin{align}
r\la v&=\frac{1}{2N_-}\left((rv+r^{\ast}v)-(rv-r^{\ast}v)\frac{\mathbf{n}\cdott\mathbf{e}}{\theta}\right)\nonumber  \\&=\frac{1}{N_-}\left(\text{Re}_{\theta}(r)v-\text{Im}_{\theta}(r)v\mathbf{n}\cdott\mathbf{e}\right)\label{eq:L action 0}\end{align}
with normalisation factor 
\begin{equation}
N_-=|\text{Re}_{\theta}(r)v-\text{Im}_{\theta}(r)v\mathbf{n}\cdott\mathbf{e}|,
\label{eq:N2}\end{equation}
where the comments made after \eqref{N} apply again. 
When $v=1+\varepsilon\mathbf{n}\cdott\mathbf{e}$, we have \begin{equation}
r\la v=1+\varepsilon\mathbf{n}\cdott\mathbf{e} ,
\end{equation}
where we have used  properties (\ref{eq:prop 2}). 
For $v=1+\varepsilon\tilde{\mathbf{m}}\cdott\mathbf{e}$, we obtain
from (\ref{eq:L action 0})  
\begin{equation}
r\la v=1+\varepsilon\mathbf{\tilde{m}}\cdott\mathbf{e}\, e^{-2\alpha}-\varepsilon\mathbf{n}\cdott\mathbf{e}\, \xi \, e^{-2\alpha} + O(\varepsilon^2).
\end{equation}

Finally, for $v=1+\varepsilon\tilde{\mathbf{n}}\cdott\mathbf{e}$
the  leading terms in the  left action are 
\begin{equation}
r\la v=1+\varepsilon\mathbf{\tilde{n}}\cdott\mathbf{e}\, e^{-2\alpha}+\varepsilon\mathbf{n}\cdott\mathbf{e}\, \eta\, e^{-2\alpha} + O(\varepsilon^2).
\end{equation}

\section{Bicrossproduct quantum groups in 3d gravity}

We now combine the general results (\ref{bi productinfini})-(\ref{bi antipode}) of Sect.~\ref{method} with the   left and right actions computed in the previous sections to obtain the Hopf algebraic structures of  the bicrossproduct quantum group  $ \C(AN) \lrbicross U(\ch)$.

\subsection{Algebra structure}
Writing simply  $\alpha$  for $\alpha \tens 1$  and similarly  $\bv n\cdott \bv e $ for  $ 1\otimes \bv n\cdott\bv e$ etc., we obtain, from (\ref{bi productinfini}), the following commutation relations in  the case  $\lambda \neq 0$:
 \begin{equation}
\begin{array}{rl}
\left[\alpha,\xi\right]&=\left[\alpha,\eta\right]=\left[\alpha,\mathbf{n}\cdott\mathbf{e}\right]=0,\\
\left[\eta,(\mathbf{m}\times\mathbf{n})\cdott\mathbf{e}\right]&=\left[\xi,\lambda\mathbf{m}\cdott\mathbf{e}\right]=0,\\
\left[\xi,\mathbf{n}\cdott\mathbf{e}\right]&=-2\lambda\eta,\\
\left[\eta,\mathbf{n}\cdott\mathbf{e})\right]&=-2\xi,\\
\left[\alpha,(\mathbf{m}\times\mathbf{n})\cdott\mathbf{e}\right]&=\pm\xi e^{-2\alpha},\\
\left[\alpha,\lambda\mathbf{m}\cdott\mathbf{e})\right]&=\mp\lambda\eta e^{-2\alpha},\\
\left[\xi,(\mathbf{m}\times\mathbf{n})\cdott\mathbf{e}\right]&=-e^{-2\alpha}\left((e^{4\alpha}-1)\mp(\xi^{2}-\lambda\eta^{2})\right),\\
\left[\eta,\lambda\mathbf{m}\cdott\mathbf{e}\right]&=-e^{-2\alpha}\left((e^{4\alpha}-1)\mp(\xi^{2}-\lambda\eta^{2})\right),\end{array}\label{alg in alpha} \end{equation}
together with the commutation relations  \eqref{comrelmnmn}.

When $\lambda=0,$ the algebra has  the commutation relation \eqref{comrelntntm} and 
\begin{equation}
\begin{array}{rl}
\left[\alpha,\xi\right]&=\left[\alpha,\eta\right]=\left[\alpha,\mathbf{n}\cdott\mathbf{e}\right]=[\xi,\tilde{\mathbf{m}}\cdott\mathbf{e}]=\left[\eta,\tilde{\mathbf{n}}\cdott\mathbf{e}\right]=0,\\
\left[\xi,\mathbf{n}\cdott\mathbf{e}\right]&=-2(e^{2\alpha}-1),\quad \left[\alpha,\tilde{\mathbf{m}}\cdott\mathbf{e}\right]=e^{-2\alpha}-1,\\
\left[\eta,\mathbf{n}\cdott\mathbf{e}\right]&=-2\xi,\quad \left[\alpha,\tilde{\mathbf{n}}\cdott\mathbf{e}\right]=\xi e^{-2\alpha},\\
\left[\xi,\tilde{\mathbf{n}}\cdott\mathbf{e}\right]&=2e^{-2\alpha}(\eta+\frac{1}{2}\xi^{2}), \quad \left[\eta,\tilde{\mathbf{m}}\cdott\mathbf{e}\right]=2(\eta+\frac{1}{2}\xi^{2})e^{-2\alpha}.\end{array}\end{equation}

\subsection{Co-algebra Structure and Antipodes}  
Next, we calculate  the co-products and antipodes.
Suppose 
\begin{equation}
r_{1}(\alpha_{1},z_{1})=(1+z_{1}Q)e^{\alpha_{1}N}\mbox{ and  }r_{2}(\alpha_{2},z_{2})=(1+z_{2}Q)e^{\alpha_{2}N},
\end{equation}
then\begin{equation}
r_{1}r_{2}=(1+(z_{1}+e^{2\alpha_{1}}z_{2})Q)e^{(\alpha_{1}+\alpha_{2})N}.\end{equation}
Hence from (\ref{Fun AN}), if we choose the continuous function $f \in \C(AN(2))$ to be  $f(r)=\alpha$,
then $
 f(r_{1}r_{2})=\alpha_{1}+\alpha_{2}$, and (\ref{bi coproduct}) gives\begin{equation}
\Delta\alpha=\alpha\otimes1+1\otimes\alpha.\end{equation}                                   
Also, with the choice of $f(r)=z$, we have 
\begin{equation}
f(r_{1}r_{2})=z_{1}+e^{2\alpha_{1}}z_{2}
\end{equation}
and (\ref{bi coproduct}) gives\begin{equation}
\Delta z=z\otimes1+e^{2\alpha}\otimes z.\end{equation}
Using the definition in (\ref{bi antipode}) we get the antipode
\begin{equation}
S(\alpha)=-\alpha\end{equation}
\begin{equation}
S(z)=-e^{-2\alpha}z.\end{equation}
  
When  $\lambda\neq0,$  we again consider $v=1+\varepsilon\mathbf{n}\cdott\mathbf{e}$. Then 
(\ref{eq:l action}) and (\ref{bi coproduct}) give the co-product
\begin{equation}
\Delta v=1\otimes1+\varepsilon(1\otimes\mathbf{n}\cdott\mathbf{e}+\mathbf{n}\cdott\mathbf{e}\otimes1) + O(\varepsilon^2),
\end{equation}
and the antipode is given by (\ref{bi antipode}) as
\begin{equation}
S(v)=1-\varepsilon\mathbf{n}\cdott\mathbf{e}.\end{equation}
When $v=1+\varepsilon(\mathbf{m}\times\mathbf{n})\cdott\mathbf{e},$
the co-product of $v$ is  given by (\ref{bi coproduct}) as
\begin{equation}
\Delta v=1\otimes1+\varepsilon(1\otimes\mathbf{(m}\times\mathbf{n})\cdott\mathbf{e}+\mathbf{(m}\times\mathbf{n})\cdott\mathbf{e}\otimes e^{-2\alpha}\pm\mathbf{n}\cdott\mathbf{e}\otimes\eta e^{-2\alpha})  + O(\varepsilon^2).
\end{equation}
The antipode is 
\begin{equation}
S(v)=1-\varepsilon(\mathbf{m}\times\mathbf{n})\cdott\mathbf{e}e^{2\alpha}\pm\varepsilon\mathbf{n}\cdott\mathbf{e}\eta + O(\varepsilon^2).
\end{equation}
 For  $v=1+\lambda\varepsilon\mathbf{m}\cdott\mathbf{e},$
 the co-product is \begin{equation}
\Delta v=1\otimes1+\varepsilon(1\otimes\lambda\mathbf{m}\cdott\mathbf{e}+\lambda\mathbf{m}\cdott\mathbf{e}\otimes e^{-2\alpha}\pm\mathbf{n}\cdott\mathbf{e}\otimes\xi e^{-2\alpha})
+ O(\varepsilon^2),
\end{equation}
and the antipode is\begin{equation}
S(v)=1-\varepsilon\lambda\mathbf{m}\cdott\mathbf{e}e^{2\alpha}\pm\varepsilon\mathbf{n}\cdott\mathbf{e}\xi + O(\varepsilon^2).\end{equation}
 
For $\lambda=0$ and $\mathbf{n}\neq0,$  we take  $v=1+\varepsilon\mathbf{n}\cdott\mathbf{e}$ and find the 
co-product
 \begin{equation}
\Delta v=1\otimes1+\varepsilon(1\otimes\mathbf{n}\cdott\mathbf{e}+\mathbf{n}\cdott\mathbf{e}\otimes1) + O(\varepsilon^2),
\end{equation}
and the antipode is given from (\ref{bi antipode}) by\begin{equation}
S(v)=1-\varepsilon\mathbf{n}\cdott\mathbf{e}.\end{equation}
When $v=1+\varepsilon\tilde{\mathbf{m}}\cdott\mathbf{e},$
the co-product is 
 \begin{equation}
\Delta v=1\otimes1+\varepsilon(1\otimes\tilde{\mathbf{m}}\cdott\mathbf{e}+\tilde{\mathbf{m}}\cdott\mathbf{e}\otimes e^{-2\alpha}-\varepsilon\mathbf{n}\cdott\mathbf{e}\otimes\xi e^{-2\alpha}) + O(\varepsilon^2),
\end{equation}
and the antipode is\begin{equation}
S(v)=1-\varepsilon\mathbf{\tilde{m}}\cdott\mathbf{e}e^{2\alpha}+\varepsilon\mathbf{n}\cdott\mathbf{e}\xi + O(\varepsilon^2).
\end{equation}
Finally, for $v=1+\varepsilon\tilde{\mathbf{n}}\cdott\mathbf{e},$
the co-product is \begin{equation}
\Delta v=1\otimes1+\varepsilon(1\otimes\tilde{\mathbf{n}}\cdott\mathbf{e}+\tilde{\mathbf{n}}\cdott\mathbf{e}\otimes e^{-2\alpha}+\mathbf{n}\cdott\mathbf{e}\otimes\eta e^{-2\alpha})
+ O(\varepsilon^2),
\end{equation}
and the antipode is \begin{equation}
S(v)=1-\varepsilon\mathbf{\tilde{n}}\cdott\mathbf{e}e^{2\alpha}+\varepsilon\mathbf{n}\cdott\mathbf{e}\eta + O(\varepsilon^2).
\end{equation}

 We now summarise the above results.  For $\lambda\neq0$, the co-products are given by
 \begin{equation}
\begin{array}{rl}
\Delta\alpha &=\alpha\otimes1+1\otimes\alpha,\\
\Delta z &=z\otimes1+e^{2\alpha}\otimes z,\\
\Delta\mathbf{n}\cdott\mathbf{e}&=1\otimes\mathbf{n}\cdott\mathbf{e}+\mathbf{n}\cdott\mathbf{e}\otimes1,\\
\Delta\mathbf{(m}\times\mathbf{n})\cdott\mathbf{e}&=1\otimes\mathbf{(m}\times\mathbf{n})\cdott\mathbf{e}+\mathbf{(m}\times\mathbf{n})\cdott\mathbf{e}\otimes e^{-2\alpha}\pm\mathbf{n}\cdott\mathbf{e}\otimes\eta e^{-2\alpha},\\
\Delta \lambda\mathbf{m}\cdott\mathbf{e}&=1\otimes \lambda\mathbf{m}\cdott\mathbf{e}+\lambda\mathbf{m}\cdott\mathbf{e}\otimes e^{-2\alpha}\mp\mathbf{n}\cdott\mathbf{e}\otimes \xi e^{-2\alpha},\end{array} 
\label{lambdanonzerocoproduct}
 \end{equation}
and the antipode is\begin{equation}
\begin{array}{rl}
S(\alpha)&=-\alpha,\\
S(z)&=-e^{-2\alpha}z,\\
S(\mathbf{n}\cdott\mathbf{e})&=-\mathbf{n}\cdott\mathbf{e},\\
S((\mathbf{m}\times\mathbf{n})\cdott\mathbf{e})&=-(\mathbf{m}\times\mathbf{n})\cdott\mathbf{e}\, e^{2\alpha}\pm\mathbf{n}\cdott\mathbf{e}\, \eta,\\
S(\lambda\mathbf{m}\cdott\mathbf{e})&=-\lambda\mathbf{m}\cdott\mathbf{e}\, e^{2\alpha}\pm\mathbf{n}\cdott\mathbf{e}\, \xi.\end{array}\label{antipodes in alpha} \end{equation}

In the case $\lambda=0,$ co-products and antipodes are given respectively by
\begin{equation}
\begin{array}{rl}
\Delta\alpha &=\alpha\otimes1+1\otimes\alpha,\\
\Delta z &=z\otimes1+e^{2\alpha}\otimes z,\\
\Delta\mathbf{n}\cdott\mathbf{e}&=1\otimes\mathbf{n}\cdott\mathbf{e}+\mathbf{n}\cdott\mathbf{e}\otimes1,\\
\Delta\tilde{\mathbf{m}}\cdott\mathbf{e}&=1\otimes\tilde{\mathbf{m}}\cdott\mathbf{e}+\tilde{\mathbf{m}}\cdott\mathbf{e}\otimes e^{-2\alpha}-\mathbf{n}\cdott\mathbf{e}\otimes\xi e^{-2\alpha},\\
\Delta\tilde{\mathbf{n}}\cdott\mathbf{e}&=1\otimes\tilde{\mathbf{n}}\cdott\mathbf{e}+\tilde{\mathbf{n}}\cdott\mathbf{e}\otimes e^{-2\alpha}+\mathbf{n}\cdott\mathbf{e}\otimes\eta e^{-2\alpha}\end{array}\end{equation}
and 
\begin{equation}
\begin{array}{rl}
S(\alpha)&=-\alpha,\\
S(z)&=-e^{-2\alpha}z,\\
S(\mathbf{n}\cdott\mathbf{e})&=-\mathbf{n}\cdott\mathbf{e},\\
S(\mathbf{\tilde{m}}\cdott\mathbf{e})&=-\mathbf{\tilde{m}}\cdott\mathbf{e}\, e^{2\alpha}+\mathbf{n}\cdott\mathbf{e}\,\xi,\\
S(\mathbf{\tilde{n}}\cdott\mathbf{e})&=-\mathbf{\tilde{n}}\cdott\mathbf{e}\,e^{2\alpha}-\mathbf{n}\cdott\mathbf{e}\, \eta.\end{array}\end{equation}

\subsection{Classical basis }
\label{classicalbasissect}
In  potential applications of the bicrossproduct $\C(AN)\lrbicross U(\ch)$ in physics (particularly in the Lorentzian case with a timelike deformation parameter, where this bicrossproduct is  the standard version of the  $\kappa$-Poincar\'e algebra), it is convenient to work in a different basis, often called the classical basis \cite{KGN2}. This basis also turns out to be convenient for studying the associated Lie bi-algebra 
and we therefore write our results in terms of classical bases, adapted to the sign of the cosmological constant. 
  
In order to match the usual conventions, we first have to replace 
$\alpha$ and $z$ by their antipodes $\tilde{\alpha}=S(\alpha)=-\alpha,$ $\tilde{z}=S(z)=-ze^{-2\alpha}$. 
The  antipode $S$ defined in (\ref{map S}) becomes 
 \begin{equation}
S(\mathbf{q})=(\tilde{w}_{3},\tilde{\xi}\mathbf{m}+\tilde{\eta}(\mathbf{m}\times\mathbf{n})+\frac{1}{\lambda}(\tilde{w}_{3}-e^{-2\tilde{\alpha}})\mathbf{n})\label{map S tilde},\end{equation}
where
\begin{equation}
\tilde{w}_{3}=\cosh2\alpha\pm\frac{1}{2}(\tilde{\xi}^{2}-\lambda\tilde{\eta}^{2})e^{-2\tilde{\alpha}}\label{ w3 tilde}\end{equation}
and $\tilde{z}=\tilde{\xi}+\theta\tilde{\eta}.$
The classical generators $\calP_{n},\calP_{mn}$ and $\calP_{m}$ are then related to the bicrossproduct basis generators according to
\begin{equation}
\begin{array}{rl}
 \calP_{m}&=\tilde{\xi}, \quad  \calP_{mn}= \tilde{\eta},\quad \calP_3=\tilde{w}_3=\lambda \calP_n+ e^{2\tilde{\alpha}}, \\[1.0ex]
\calP_{n}&=\frac{1}{\lambda}\left( \sinh2\tilde{\alpha} \pm\frac{  1}{  2}(\tilde{\xi}^{2}-\lambda\tilde{\eta}^{2})e^{-2\tilde{\alpha}}\right),
\end{array}
\end{equation}
and satisfy the constraint
\begin{equation}
\label{classicalconstraint}
 \calP_{3}^{2}\mp \calP_m^2 \pm \lambda \calP_{mn}^{2}-\lambda^2\calP_{n}^{2}=1.\end{equation}
From these, we obtain the algebra
\begin{equation}
\begin{array}{rl}
\left[\calP_{mn},(\mathbf{m}\times\mathbf{n})\cdott\mathbf{e}\right]&=\left[\calP_{m},\lambda\mathbf{m}\cdott\mathbf{e}\right]=\left[\calP_{n},\mathbf{n}\cdott\mathbf{e}\right]=0,\\
\left[\calP_{m},\mathbf{n}\cdott\mathbf{e}\right]&=-2\lambda \calP_{mn},\quad\left[\calP_{mn},\mathbf{n}\cdott\mathbf{e}\right]=-2\calP_{m},\\
\left[\calP_{n},(\mathbf{m}\times\mathbf{n})\cdott\mathbf{e}\right]&=\mp2 \calP_{m},\quad\left[\calP_{m},(\mathbf{m}\times\mathbf{n})\cdott\mathbf{e}\right]=2\calP_{n},\\
\left[\calP_{n},\lambda\mathbf{m}\cdott\mathbf{e}\right]&=\pm2\lambda \calP_{mn},\quad\left[\calP_{mn},\lambda\mathbf{m}\cdott\mathbf{e}\right]=2 \calP_{n}.\end{array}
\label{classicalone}
\end{equation}
together with the commutators \eqref{comrelmnmn}. 

Defining the vector 
\begin{equation}
\label{Pdeff}
 \calP = \calP_n \bn  +  \calP_{m} \bm +  \calP_{mn} \bm\times\bn,
\end{equation}
we have 
\begin{equation}
 \bn\cdott \calP = -\lambda \calP_n,\quad \bm\cdott \calP = \mp \frac{1}{\lambda} \calP_m,\quad (\bm\times \bn) \cdott \calP = \pm  \calP_{mn}.
\end{equation}
In terms of the components $\calP_a$, $a=0,1,2$ of $\calP$ with respect to any orthonormal basis, and the generators $J_a=\frac 1 2 e_a$, the  brackets  \eqref{classicalone} become the `classical' brackets
\begin{equation}
\label{classicaltwo}
 [J_a,J_b]=\epsilon_{abc}J^c, \quad [J_a,\calP_b]=\epsilon_{abc}\calP^c,\quad [\calP_a,\calP_b]=0,
\end{equation}
and the constraint \eqref{classicalconstraint} is simply 
\begin{equation}
\label{classicalconstraintt}
 \calP_3^2 + \lambda \calP^2 =1.
\end{equation}

In order to write down the co-algebra, we introduce
\begin{equation} 
\vec{\calP}^{2}=\calP_{m}^{2}-\lambda \calP_{mn}^{2}, \quad  \text{and} \quad 
 T=e^{2\tilde{\alpha}}=\left[1+\lambda^2 \calP_{n}^{2}\pm  \vec{\calP}^2 \right]^{\frac{1}{2}}-\lambda \calP_{n}.
\end{equation}
 Then the 
 co-products of the classical generators are given by  
  \begin{equation}
\begin{array}{rl}
\Delta \calP_{n}&=\frac{  1}{  2\lambda}\left(T\otimes T-T^{-1}\otimes T^{-1}\right)\\[1.0ex]
&\pm\frac{  1}{  2\lambda}\left(T^{-1}\vec{\calP}^{2}\otimes T+2T^{-1} \left(\calP_{m}\otimes \calP_{m}-\lambda \calP_{mn}\otimes \calP_{mn}\right)+T^{-1}\otimes T^{-1} \vec{\calP}^{2}\right),\\[1.0ex]
\Delta \calP_{m}&=\calP_{m}\otimes T+1\otimes \calP_{m},\\
\Delta \calP_{mn}&=\calP_{mn}\otimes T+ 1\otimes \calP_{mn},\\
\Delta\mathbf{n}\cdott\mathbf{e}&=1\otimes\mathbf{n}\cdott\mathbf{e}+\mathbf{n}\cdott\mathbf{e}\otimes1,\\
\Delta\mathbf{(m}\times\mathbf{n})\cdott\mathbf{e}&=1\otimes\mathbf{(m}\times\mathbf{n})\cdott\mathbf{e}+\mathbf{(m}\times\mathbf{n})\cdott\mathbf{e}\otimes T \pm\mathbf{n}\cdott\mathbf{e}\otimes \calP_{mn},\\
\Delta \lambda\mathbf{m}\cdott\mathbf{e}&=1\otimes \lambda\mathbf{m}\cdott\mathbf{e}+\lambda\mathbf{m}\cdott\mathbf{e}\otimes T \mp\mathbf{n}\cdott\mathbf{e}\otimes \calP_{m}.
\end{array}
\label{classicalcoproductnonzero}
\end{equation}

When  $\lambda=0,$ the  classical basis   is  related to the bicrossproduct
basis by 
\begin{equation}
\begin{array}{rl}
\mathcal{P}_{n} &=e^{-2\alpha}\left(\eta+\frac{1}{2}\xi^{2}\right),\\[1.0ex]
\mathcal{P}_{\tilde{m}}&=\xi \quad \mathcal{P}_{\tilde{n}}=e^{2\alpha}-1.
\end{array}
\end{equation}
In analogy to \eqref{Pdeff}, we define the vector 
\begin{equation}
\label{Pdef1}
 \calP = \calP_n \bn  +  \calP_{\tilde{m}} \tilde{\bm} +  \calP_{\tilde{n}} \tilde{\bn},
\end{equation}
so that 
\begin{equation}
 \bn\cdott \calP =  \calP_{\tilde{n}},\quad \tilde{\bm}\cdott \calP = - \calP_{\tilde{m}},\quad \tilde{\bn}\cdott \calP = \calP_{n}.
\end{equation}
In this basis, we obtain the Lie  algebra
\begin{equation}
\begin{array}{rl}\left[\mathcal{P}_{\tilde{n}},\mathbf{n}\cdott\mathbf{e}\right]&=\left[\mathcal{P}_{\tilde{m}},\tilde{\mathbf{m}}\cdott\mathbf{e}\right]=\left[\mathcal{P}_{n},\tilde{\mathbf{n}}\cdott\mathbf{e}\right]=0,\\
\left[\mathcal{P}_{n},\mathbf{n}\cdott\mathbf{e}\right]&=-2\mathcal{P}_{\tilde{m}},\quad\left[\mathcal{P}_{\tilde{m}},\mathbf{n}\cdott\mathbf{e}\right]=-2\mathcal{P}_{\tilde{n}},\\
\left[\mathcal{P}_{\tilde{n}},\mathbf{\tilde{n}}\cdott\mathbf{e}\right]&=2\mathcal{P}_{\tilde{m}},\quad\left[\mathcal{P}_{n},\tilde{\mathbf{m}}\cdott\mathbf{e}\right]=2\mathcal{P}_{n},\\
\left[\mathcal{P}_{\tilde{n}},\mathbf{\tilde{m}}\cdott\mathbf{e}\right]&=-2\mathcal{P}_{\tilde{n}},\quad\left[\mathcal{P}_{\tilde{m}},\mathbf{\tilde{n}}\cdott\mathbf{e}\right]=2\mathcal{P}_{n},
\end{array}\end{equation}
together with \eqref{comrelntntm},
which is again equivalent to \eqref{classicaltwo}. 

With $\tilde{T}=e^{2\alpha}=1+\CP_{\tilde{n}} = 1+\bn\cdott\CP$, 
the  co-products are 
\begin{equation}
\begin{array}{rl}
\Delta \CP_{n}&=1\otimes \CP_{n}+\CP_{n}\otimes\tilde{T}^{-1}+\frac{1}{2}\CP_{\tilde{n}}\otimes\tilde{T}^{-1}\CP_{\tilde{m}}^{2}+\CP_{\tilde{m}}\otimes\tilde{T}^{-1}\CP_{\tilde{m}},\\
\Delta \CP_{\tilde{m}}&=\CP_{\tilde{m}}\otimes1+\tilde{T}\otimes \CP_{\tilde{m}},\\
\Delta \CP_{\tilde{n}}&=\CP_{\tilde{n}}\otimes \CP_{\tilde{n}},\\
\Delta\mathbf{n}\cdott\mathbf{e}&=1\otimes\mathbf{n}\cdott\mathbf{e}+\mathbf{n}\cdott\mathbf{e}\otimes1,\\
\Delta\tilde{\mathbf{m}}\cdott\mathbf{e}&=1\otimes\tilde{\mathbf{m}}\cdott\mathbf{e}+\tilde{\mathbf{m}}\cdott\mathbf{e}\otimes\tilde{T}^{-1}-\mathbf{n}\cdott\mathbf{e}\otimes\tilde{T}^{-1}\CP_{\tilde{m}},\\
\Delta\tilde{\mathbf{n}}\cdott\mathbf{e}&=1\otimes\tilde{\mathbf{n}}\cdott\mathbf{e}+\tilde{\mathbf{n}}\cdott\mathbf{e}\otimes\tilde{T}^{-1}+\mathbf{n}\cdott\mathbf{e}\otimes \CP_{n}-\frac{1}{2}\mathbf{n}\cdott\mathbf{e}\otimes\tilde{T}^{-1}\CP_{\tilde{m}}^{2}. \end{array}
\label{classicalcoproductzero}
\end{equation}

\subsection{Summary of results}

 In Table~\ref{resulttable}
we list the  semiduals  computed in this paper of the universal enveloping algebras of the Lie algebras for the groups and factorisation given in  Table~\ref{doublecrosstable}. 
In the Euclidean case with $\lambda>0,$  semidualisation  of the enveloping algebra
$U(\mathfrak{su}(2)\rcross \mathfrak{su}(2))$ 
 gives the quantum double $D(U(\mathfrak{su}(2)))=\C(SU(2)) \lcross U(\mathfrak{su}(2)) $.
For  $\lambda=0$ and Euclidean signature,  the  semidual of $U(\mathfrak{su}(2)\rcross \R^3)$ is $(\R^*)^3 \lcross  U(\mathfrak{su}(2)) .$  In the Lorentzian case and  with $\lambda>0$,
semidualisation of  $U(\mathfrak{sl}(2,\R) \rcross U(\mathfrak{sl}(2,\R))$ gives  the quantum double $D(U(\mathfrak{sl}(2,\R)))= \C(SL(2,\R))\lcross  U(\mathfrak{sl}(2,\R))$. When $\lambda=0$, 
the semidual  of  $U(\mathfrak{sl}(2,\R )\rcross \R^3) $  is   $(\R^*)^3  \lcross  U(\mathfrak{sl}(2,\R ))$.  All this follows from our discussion of examples
in Sect.~\ref{simpleexamples}. 
We do not give details of the (standard) quantum group structure for these cases, but see  \cite{MajidSchroers} for details of the Euclidean case. 
For  the (non-trivial) bicrossproducts  in Table~\ref{resulttable}  we use the notation $\lrbicross_s,$ $\lrbicross_l$ and $\lrbicross_t$ for the  left-right bicrossproducts with spacelike, lightlike and timelike deformations respectively. 
In order to make contact with the standard  literature on $\kappa$-Poincar\'e algebras and related bicrossproducts we   discuss  each of these  in conventional notation in the Appendix.    The first treatments  of $\kappa$-Poincar\'e symmetry focused on the Lorentzian case, with a timelike deformation vector \cite{LNRT,LNR} but spacelike \cite{BHDSspacelikedeformation} and lightlike \cite{BHDSlightlikedeformation} deformation vectors were soon considered.   In the   Euclidean case different choices of $\bn$ lead to isomorphic Lie bialgebras, see e.g. Chapter 8 of  \cite{Majidbook1} for a details.

%

\begin{table}[H]
\centering
\begin{tabular}{|l|c|c|}
\hline
&&\\
 & Euclidean signature  & Lorentzian signature\\
&&\\
\hline
&&\\
$ \lambda>0$ & $D(U(\mathfrak{su}(2))) $ & $ D(U(\mathfrak{sl}(2,\R)))$ \\                                                                                                                                                           
                        &                        & $\C(AN(2))\lrbicross_s U(\mathfrak{sl}(2,\R))  $     \\          
&&\\
\hline
&&\\
$ \lambda=0$ & $ (\R^*)^3  \lcross  U(\mathfrak{su}(2))
$ & $(\R^*)^3  \lcross  U(\mathfrak{sl}(2,\R ))$    \\
             &                               & $  \C(AN(2))\lrbicross_l    U(\mathfrak{sl}(2,\R))$\\
&&\\
\hline
&&\\
$ \lambda<0$ & $\C(AN(2))\lrbicross U(\mathfrak{su}(2))   $& $  \C(AN(2)) \lrbicross_t U(\mathfrak{sl}(2,\R))$\\
&&\\

\hline
\end{tabular}

\caption{Semiduals of local isometry groups in 3d gravity}
\label{resulttable}
 \end{table}

\section{Discussion and conclusion}
\subsection{Classical $r$-matrices}
\label{rmatrixsect}
Before we discuss our results we  consider the Lie bi-algebra structures for each of the bicrossproduct Hopf algebras we have obtained. 
We work in the classical basis of Sect.~\ref{classicalbasissect} where the Lie brackets have the form \eqref{classicaltwo}. This shows  in particular that, for all values of the cosmological constant, the Lie algebra is $\cg= \RR^3\lcross  \ch$, which is the Lie algebra of the Euclidean group in the Euclidean case, and the Lie algebra of the Poincar\'e group in the Lorentzian case. 

In order to extract  the co-commutator for a generator $Y$, we 
compute $\Delta(Y)-\Delta^{op}(Y)$ for each of the  co-products in 
\eqref{classicalcoproductnonzero} for $\lambda \neq 0$ and  in \eqref{classicalcoproductzero} for $\lambda =0$,  and keep  only leading (quadratic) terms.
For $\lambda\neq 0$, we find  
 \begin{equation}
\begin{array}{rl}
\delta(\bn\cdott \calP)&=\delta(\mathbf{n}\cdott\mathbf{e})=0,\\
\delta(\bm\cdott \calP)&=\bn\cdott \calP\wedge \bm\cdott\calP,\\
\delta((\bm\times \bn)\cdott \calP&= \bn\cdott \calP\wedge (\bm\times\bn)\cdott\calP,\\
\delta(\mathbf{m}\cdott\mathbf{e})&=\bn\cdott \calP\wedge\bm\cdott\mathbf{e}
+\bn\cdott\mathbf{e}\wedge \bm \cdott \calP,\\
\delta((\mathbf{m}\times\mathbf{n})\cdott\mathbf{e})&=\bn\cdott \calP\wedge (\bm\times\bn)\cdott\mathbf{e}+\bn\cdott\mathbf{e}\wedge (\bm \times \bn) \cdott \calP,\\
\end{array}
\label{cocomlambdanonzero}
\end{equation}
with $\wedge$ denoting the skewsymmetric tensor product.
For $\lambda = 0$, we obtain
 \begin{equation}
\begin{array}{rl}
\delta(\bn\cdott \calP)&=\delta(\mathbf{n}\cdott\mathbf{e})=0,\\
\delta(\tilde{\bn}\cdott \calP)&=\bn\cdott \calP\wedge \tilde{\bn}\cdott\calP,\\
\delta(\tilde{\bm}\cdott \calP&= \bn\cdott \calP\wedge \tilde{\bm}\cdott\calP,\\
\delta(\tilde{\bn}\cdott\mathbf{e})&=\bn\cdott \calP\wedge\tilde{\bn}\cdott\mathbf{e}
+\bn\cdott\mathbf{e}\wedge \tilde{\bn}\cdott \calP,\\
\delta(\tilde{\bm} \cdott\mathbf{e})&=\bn\cdott \calP\wedge \tilde{\bm}\cdott\mathbf{e}+\bn\cdott\mathbf{e}\wedge \tilde{\bm}  \cdott \calP.\\
\end{array}
\label{cocomlambdazero}
\end{equation}

A co-commutator for a Lie algebra $\cg$  is co-boundary if its action on any $Y\in \cg$  can be written as  \begin{equation}
\delta(Y)=\left[1\otimes Y+Y\otimes1,r\right],\end{equation}
in terms of an element $r\in \cg\otimes \cg$, called the classical $r$-matrix.
Both the co-commutators \eqref{cocomlambdanonzero} and \eqref{cocomlambdazero} turn out to be co-boundary. The $r$-matrix  in both cases is  \cite{MeusburgerSchroers7}
\begin{equation}
\label{rmatrix}
r_{\bn}=-\epsilon_{abc}n^c J^a\wedge \calP^b.
\end{equation}
 This is the classical $r$-matrix of  the 3d $\kappa$-Poincar\'e (or Euclidean)  algebra with  
deformation vector $\bn$.

For completeness we also note the classical $r$-matrices for the quantum doubles $D(U(\mathfrak{su}(2))) $  and  $ D(U(\mathfrak{sl}(2,\R)))$ in Table~\eqref{resulttable}, which arise when $\lambda >0$. These are well-known \cite{Schroers,MajidSchroers} and easily computed.  In the semiclassical basis used   in  ~\cite{MajidSchroers}  one deduces the co-product from the multiplication rule of  quaternions of the form $P_0 + \sqrt{\lambda}\, \be \cdot \calP$. Extracting leading terms one finds the co-commutator
\begin{equation}
 \delta(\calP_a) =\sqrt{\lambda}\,\epsilon_{abc} \,\calP^b\wedge\calP^c,
\end{equation}
valid in both the Euclidean and Lorentzian setting. This is co-boundary, with classical $r$-matrix
\begin{equation}
\label{doublermatrix}
 r_D= \sqrt{\lambda}\, \calP_a\wedge J^a,
\end{equation}
where we have omitted symmetric, invariant terms.

\subsection{Interpretation in the context of 3d gravity}


We have seen that the  algebra structure of the  Hopf algebra we obtain via semidualisation  is independent of the cosmological parameter $\lambda$ and  always that of a  universal enveloping algebra:  that of the Euclidean Lie algebra in the Euclidean case, and that of the Poincar\'e Lie algebra in the Lorentzian case.  However, the co-algebra structure  depends both on the  original isometry group (i.e. on $\lambda$) and  on the factorisation (i.e. on the vector $\bn$ in the cases where it is defined). 

In interpreting our results it is important to keep in mind the distinction between the original isometry groups  with the Lie algebras $\mathfrak{g}_\lambda$ generated by $J_a$ and $P_a$, $a=0,1,2$, and the semidual Lie  brackets \eqref{classicaltwo} of the Euclidean or Poincar\'e Lie algebra with generators $J_a$ and $\calP_a$, $a=0,1,2$.
The `dual momenta' $\calP_a$  generate the (abelian) algebra dual to that of the original momentum algebra $U(\mathbf{an}(2)_{\bn})$  
with generators $\tilde{P}_a$, $a=0,1,2$ \eqref{Iwa decomp}. From the point of view of the original symmetry algebras one should really  think of the $\calP_a$ as coordinates on position space: the constraint \eqref{classicalconstraintt} is precisely the constraint \eqref{constraint} defining the model spacetimes. However, in keeping with the philosophy and terminology of \cite{MajidSchroers} we combine the mathematical operation of semidualisation with a shift of interpretation: we think of the $\calP_a$ as momentum generators in a dual theory, and ask how the applicability regimes of the  original and dual theory are related. In this way, our interpretation of  the generators $\calP_a$ also agrees with that  in the physics literature on   $\kappa$-Poincar\'e or quantum double symmetry.

In order to  characterise and distinguish the original and the semidual theory, it is worth recalling  \cite{sissatalk} that in 3d gravity 
the curvature  of the model spacetime  is controlled by the cosmological constant, with  `curvature radius' 
\begin{equation}
\label{cosmolength}
 \tau_C=\frac{1}{\sqrt{\lambda}},
\end{equation}
where we used the letter $\tau_C$ to remind the reader that  $\tau_C$ has the physical dimension of time, since  the cosmological constant $\Lambda_C$ has physical dimension of inverse length squared and  $\lambda =-c^2\Lambda_C$.
The curvature of momentum space, on the other hand,  is controlled by Newton's constant  $G$,   with the curvature radius given by the 3d Planck mass:
\begin{equation}
 \mu_p=\frac{1}{16\pi G}.
\end{equation}
Note that our  definition of the Planck mass differs by a factor of two from the one which is most commonly used in 3d gravity; we adopt it here merely to simplify expressions in the following general discussion

The above   summary is a useful guideline even though it is  strictly speaking only true when either the cosmological constant or Newton's constant is zero; in general the two parameters combine into a new dimensionless parameter which affects both spacetime and momentum curvature
\cite{SchroersCracow}.  As further discussed  in \cite{MajidSchroers} and with the same caveat, 
the cosmological constant controls the algebra structure, while the Planck mass controls the co-algebra structure in the quantum isometry groups of 3d quantum gravity. 
One should thus think of the universal enveloping algebras $U(\mathfrak{g}_\lambda)$ as 
quantum isometry groups  in a regime with flat momentum space  (and hence infinite Planck mass) and curved spacetime,  controlled by the cosmological parameter   $\lambda$. 
 By contrast, the semidual quantum groups are associated to flat model spacetimes (since the  `semidual  cosmological constant' is zero in the brackets \eqref{classicaltwo}) but curved momentum space (since the co-algebra structures are non-trivial).

We now want to relate the cosmological parameter $\lambda$  of the original theory  to  the Planck mass $\tilde{\mu}_P$  in the semidual theory. 
To do this, we   briefly recall aspects of  the combinatorial quantisation programme which we  touched on  in the introduction.  This programme is based on the Chern-Simons formulation of 3d gravity, where the local isometry groups  play the role of gauge groups. The Chern-Simons action requires an invariant, non-degenerate inner product on the Lie algebra  of the relevant local isometry group. 
In order to write  the Einstein-Hilbert action as a  Chern-Simons action (up to boundary terms) for the (semidual) Lie algebra spanned by $J_a,\calP_a$, $a=1,2,3$,   one  needs to take the inner product   with non-vanishing pairings
\begin{equation}
\label{inprod}
 \langle J_a, \calP_b\rangle =\frac{1}{16 \pi \tilde{G}}\;\eta_{ab},
\end{equation}
where $\tilde{G}$ is Newton's constant in the semidual theory,  again related to the Planck mass via $\tilde{\mu}_p=1/(16\pi \tilde{G})$.

The combinatorial quantisation programme is based on a description of the Poisson structure on an extended phase space, due to Fock and Rosly \cite{FockRosly}. The latter  makes essential use of a classical $r$-matrix which is required to be compatible with the inner product \eqref{inprod} in the sense that it satisfies the classical Yang-Baxter equation and  its symmetric part  is equal to the Casimir
\begin{equation}
\label{casimir}
 K= \frac{1}{\tilde{\mu}_P} \left( J_a \otimes  \calP^a +\calP_a\otimes J^a\right)
\end{equation}
associated  to \eqref{inprod}. 

All the $r$-matrices we computed in  Sect.~\ref{rmatrixsect} are antisymmetric. To ascertain  their compatibility with 3d gravity we therefore need to check if their sum with  the Casimir \eqref{casimir} satisfies the classical Yang-Baxter equation. This was systematically investigated in \cite{MeusburgerSchroers7}. The result is that the  $r$-matrix $r_{\bn}$  \eqref{rmatrix} of bicrossproduct type
  is compatible  with the inner product  \eqref{inprod} iff 
\begin{equation}
 \mathbf{n}^2= - \frac{1}{\tilde{\mu}_P^2}.
\end{equation}
 Using  $\mathbf{n}^2=-\lambda$ we deduce the condition 
\begin{equation}
\label{mastercondition}
\lambda =  \frac{1} {\tilde{\mu}^2_P}. 
\end{equation}
Similarly, the $r$-matrix \eqref{doublermatrix}
of the doubles $D(U(\mathfrak{su}(2))) $  and  $ D(U(\mathfrak{sl}(2,\R)))$  is  compatible with the Casimir \eqref{casimir} if $\sqrt{\lambda}= 1/\tilde{\mu}_P$. Since   $\lambda>0$ by assumption in this case,  this requirement  is equivalent to the condition \eqref{mastercondition}, which therefore covers all cases.

The  schematic summary of  our results in Table~\ref{semidualtable} shows that semidualisation may be viewed as  an exchange $
 \mu_P \leftrightarrow {\tau_C}$, 
confirming and 
extending  the result obtained in the  Euclidean setting in \cite{MajidSchroers}.  
The mismatch of physical dimensions in this exchange  is a consequence of the exchange of position and momentum degrees of freedom under the semidualisation map.  In the models considered here (where either the cosmological time scale or the Planck mass is infinite) this has to be repaired `by hand',  by multiplying with a suitable dimensionful constant. To understand this better one needs to go to a regime where {\em both} the cosmological time scale and the Planck mass have finite values. The theory in that regime is controlled by a dimensionless parameter, essentially an exponential of $\hbar/(\mu_P \tau_C)$,  which is manifestly invariant under the exchange of cosmological time scale and Planck mass \cite{MajidSchroers}. The limits $\mu_P\rightarrow \infty$ and $\tau_C\rightarrow \infty$ correspond to different contractions of the associated quantum isometry groups. The contractions   require the introduction of dimensionful constants, and this is the origin of the different physical dimensions of the remaining parameter after contraction.

\begin{table}[H]
\centering
\begin{tabular}{|c|c|c|}
\hline
&&\\
 & Original   regime  & Semidual regime \\
&&\\
\hline
&&\\
 Cosmological &   $\tau_C=\frac{1}{\sqrt{\lambda}}$  & $ \tilde{\tau}_C= \infty $\\                                                                                                                                                           
                          time scale &&\\
\hline
&&\\
Planck mass  &  $\mu_P= \infty $& $\tilde{\mu}_P= \tau_C$    \\
&&\\
\hline
\end{tabular}

\caption{Semiduality of regimes in 3d gravity.}
\label{semidualtable}
 \end{table}
While  the parameter $\lambda$ naturally takes any real  value,  a physically reasonable Planck mass should be real.  At first sight it is therefore a concern that the  condition \eqref{mastercondition} may force the  semidual Planck mass to be imaginary.  However, formally extending our interpretation of  3d gravity  to include imaginary Planck mass   we  obtain  a very symmetric picture, and  a gravitational interpretation of all quantum groups constructed in this paper  via semiduality.  In particular, one could then rephrase the result of \cite{MeusburgerSchroers7} that the  $\kappa$-Poincar\'e algebra with a timelike deformation vector $\bn$ is not associated to 3d quantum gravity by saying that the $\kappa$-Poincar\'e algebra with a timelike deformation vector $\bn$ describes 3d quantum gravity with an imaginary Planck mass.

Finally, we note  that pairs of  bicrossproducts which arise, via semiduality,  from two different factorisations of the same isometry group  have twist-equivalent associated Lie bialgebras. This follows from the fact that they are  equivalent as algebras and that their associated classical $r$-matrices are compatible with the same inner product \eqref{inprod}.  The most interesting example of such a pair of bicrossproducts is the quantum double $ D(U(\mathfrak{sl}(2,\R)))$  and the bicrossproduct $\C(AN(2))\lrbicross_s U(\mathfrak{sl}(2,\R))  $  with a spacelike deformation parameter, which are therefore equally valid quantum isometry groups in the combinatorial factorisation programme. Here  the consideration of their semiduals provides  a link  between two quantum groups whose relation would otherwise be somewhat mysterious. 

\vspace{0.4cm}

\noindent {\bf Acknowledgments} PKO acknowledges an ICTP PhD  fellowship and a UG Carnegie Corporation fund  supporting extended annual visits to  Heriot-Watt University. BJS and PKO thank the Edinburgh Mathematical Society for additional support and  gratefully acknowledge the hospitality of AIMS South Africa and the Perimeter Institute, where part of this work was carried out.

\appendix

\section{The results in conventional notations}

The quaternionic formalism used in this paper is very effective for our purposes but leads to a description of bicrossproduct Hopf algebras like the $\kappa$-Poincar\'e and $\kappa$-Euclidean algebras which looks rather different from that given in the standard  literature.  In this appendix 
we therefore summarise the algebra and co-algebra structure of each of the bicrossproduct Hopf algebras in Table~\ref{resulttable}  in conventional notation.

\subsection{The case $\lambda \neq 0$}

For the case $\lambda<0$ and $c^2<0$ (Euclidean signature),  the semidual Hopf algebra in Table~\ref{resulttable} is  the Euclidean bicrossproduct  $\C(AN(2)) \lrbicross U(\mathfrak{su}(2))$.
The translation generators $p_a, a=0,1,2,$ rotation $M$ and boosts $N_i, i=1,2,$  are given by 
\begin{equation}
\begin{array}{rl}
2\alpha &=\frac{  p_0}{  \sqrt{-\lambda}} ,\quad \xi=\frac{  p_2}{  \sqrt{-\lambda}}, \quad \eta=\frac{  p_1}{  \lambda}, \\
\mbox{and } M &=\frac{  1}{  2\sqrt{-\lambda}}\bv n \cdott \bv e,\quad N_1=-\frac{  1}{  2}(\bv m \times \bv n)\cdott \bv e, \quad  N_2=\frac{  \sqrt{-\lambda}}{  2}\bv m \cdott \bv e. \end{array}\end{equation} 
This gives the algebra
\begin{equation}
\begin{array}{rl}
\left[p_a,p_b\right]&=0,\quad [M,N_1]=-N_2, \quad [M,N_2]=N_1, \quad [N_1,N_2]=-M,\\[1.0ex]
\left[p_0,M\right]&=0 \quad \left[p_i,M\right]=\epsilon_{ij} p_j,   \\[1.0ex]
\left[p_0, N_i\right]&=-\epsilon_{ij} p_j e^{-\frac{p_0}{\sqrt{-\lambda}}},   \\[1.0ex]
\left[p_i,N_j\right]&=-\epsilon_{ij}e^{-\frac{p_0}{\sqrt{-\lambda}}}\left(\frac{ \sqrt{-\lambda}}{ 2}(e^{\frac{2p_0}{\sqrt{-\lambda}}}-1)- \frac{1}{ 2\sqrt{-\lambda}}\vec{ p}^2 \right), \quad i,j=1,2,\\[1.5ex]
\end{array} \end{equation}
where $\vec{p}^2=p_1^2+p_2^2.$ The co-products are 
 \begin{equation}
\begin{array}{rl}
\Delta p_0 &=p_0\otimes1+1\otimes p_0,\\[1.0ex]
\Delta p_i &=p_i\otimes1+e^{\frac{p_0}{\sqrt{-\lambda}}}\otimes p_i,\\[1.0ex]
\Delta M&=1\otimes M+M\otimes 1,\\[1.0ex]
\Delta N_i&=1\otimes N_i+N_i \otimes e^{-\frac{p_0}{\sqrt{-\lambda}}}+\frac{1}{ \sqrt{-\lambda}}M\tens p_i e^{-\frac{p_0}{\sqrt{-\lambda}}} ,\quad i=1,2. \end{array}  \end{equation}
One can match the conventions of the algebra described in \cite{MajidRuegg}    by choosing basis in which 
\begin{equation}
\tilde{p}_i=p_ie^{-\frac{p_0}{\sqrt{-\lambda}}}.                                            
\end{equation}
In this case, we obtain the algebra
\begin{equation}
\begin{array}{rl}
\left[\tilde{p}_a,\tilde{p}_b\right] &=0, \quad [M,N_1]=-N_2, \quad [M,N_2]=N_1, \quad [N_1,N_2]=-M,\\[1.0ex]
\left[\tilde{p}_0,M\right]&=0 \quad \left[\tilde{p}_i,M\right]=\epsilon_{ij}\tilde{p}_j,\\[1.0ex]
\left[p_0, N_i\right]&=-\epsilon_{ij}\tilde{p}_j,\quad i,j=1,2, \\[1.0ex]
\left[\tilde{p}_1,N_1\right]&=\frac{1}{\sqrt{-\lambda}} \tilde{p}_1\tilde{p}_2, \quad [\tilde{p}_2,N_2]=-\frac{1}{\sqrt{-\lambda}}\tilde{p}_1\tilde{p}_2,\\[1.0ex]
\left[\tilde{p}_1,N_2\right]&=-\frac{ \sqrt{-\lambda}}{ 2}(1-e^{-\frac{2p_0}{\sqrt{-\lambda}}})+ \frac{ 1}{2\sqrt{-\lambda}}(\tilde{p}_2^2-\tilde{p}_1^2 ),\\[1.5ex]
\left[\tilde{p}_2,N_1\right]&=\frac{ \sqrt{-\lambda}}{ 2}(1-e^{-\frac{2p_0}{\sqrt{-\lambda}}})+ \frac{1}{ 2\sqrt{-\lambda}}(\tilde{p}_2^2-\tilde{p}_1^2 ).\\[1.5ex]
\end{array} \end{equation}
with co-products 
 \begin{equation}
\begin{array}{rl}
\Delta p_0 &=p_0\otimes1+1\otimes p_0,\\[1.0ex]
\Delta \tilde{p}_i &=\tilde{p}_i\otimes e^{-\frac{p_0}{\sqrt{-\lambda}}}+ 1\otimes \tilde{p}_i,\quad i=1,2\\[1.0ex]
\Delta M&=1\otimes M+M\otimes 1,\\[1.0ex]
\Delta N_i&=1\otimes N_i+N_i \otimes e^{-\frac{p_0}{\sqrt{-\lambda}}}+ \frac{1}{ \sqrt{-\lambda}} M\tens \tilde{p}_i,\quad i=1,2. \end{array}  \end{equation}

When $\lambda<0$  and $c^2>0$  (Lorentzian signature),  the bicrossproduct in  Table~\ref{resulttable}  $ \C(AN(2))\lrbicross_t U(\mathfrak{sl}(2,\R))$ ($\kappa$-Poincar\'e algebra with a  timelike deformation parameter).
Setting 
\begin{equation}
\begin{array}{rl}
2\alpha &=\frac{  p_0}{  \sqrt{-\lambda}} ,\quad \xi=\frac{  p_2}{  \sqrt{-\lambda}}, \quad \eta=\frac{  p_1}{  \lambda}, \\
\mbox{and } M &=\frac{  1}{  2\sqrt{-\lambda}}\bv n \cdott \bv e,\quad N_1=-\frac{  1}{  2}(\bv m \times \bv n)\cdott \bv e, \quad  N_2=\frac{  \sqrt{-\lambda}}{  2}\bv m \cdott \bv e, \end{array}  \end{equation} 
we  have the algebra
\begin{equation}
\begin{array}{rl}
\left[p_a,p_b\right] &=0,\quad [M,N_1]=-N_2, \quad [M,N_2]=N_1, \quad [N_1,N_2]=-M, \\[1.0ex]
\left[p_0,M\right]&=0 \quad \left[p_i,M\right]=\epsilon_{ij} p_j,\quad i,j=1,2 \\[1.0ex]
\left[p_0, N_i\right]&=-\epsilon_{ij}p_j e^{-\frac{p_0}{\sqrt{-\lambda}}}, \quad i=1,2 , \\[1.0ex]
\left[p_i,N_j\right]&=-\epsilon_{ij}e^{-\frac{p_0}{\sqrt{-\lambda}}}\left(\frac{ \sqrt{-\lambda}}{ 2}(e^{\frac{2p_0}{\sqrt{-\lambda}}}-1)+ \frac{1}{ 2\sqrt{-\lambda}}\vec{p}^2 \right),  \quad i=1,2,\\[1.5ex]
\end{array} \end{equation}
where $\vec{p}^2 = p_1^2+p_2^2.$ The co-products are 
 \begin{equation}
\begin{array}{rl}
\Delta p_0 &=p_0\otimes1+1\otimes p_0,\\[1.0ex]
\Delta p_i &=p_i\otimes1+e^{\frac{p_0}{\sqrt{-\lambda}}}\otimes p_i, \\[1.0ex]
\Delta M&=1\otimes M+M\otimes 1,\\[1.0ex]
\Delta N_i&=1\otimes N_i+N_i \otimes e^{-\frac{p_0}{\sqrt{-\lambda}}}- \frac{1}{ \sqrt{-\lambda}}M \tens p_i e^{-\frac{p_0}{\sqrt{-\lambda}}} ,\quad i=1,2. \end{array}  \end{equation}

For the case $\lambda>0$ and $c^2>0$ (Lorentzian signature),  the bicrossproduct in Table~\ref{resulttable} is  $\C(AN(2))\lrbicross_s U(\mathfrak{sl}(2,\R))$ ($\kappa$-Poincar\'e algebra with a spacelike deformation parameter).
The translation generators $p_a, a=0,1,2,$ rotation $M$ and boosts $N_i, i=1,2,$  are given by 
\begin{equation}
\begin{array}{rl}
2\alpha &=\frac{  p_1}{  \sqrt{\lambda}} ,\quad \xi=\frac{  p_2}{  \sqrt{\lambda}}, \quad \eta=\frac{  p_0}{  \lambda}, \\
\mbox{and } N_1 &=\frac{  1}{  2\sqrt{\lambda}}\bv n \cdott \bv e,\quad M=-\frac{  1}{  2}(\bv m \times \bv n)\cdott \bv e, \quad  N_2=\frac{  \sqrt{\lambda}}{  2}\bv m \cdott \bv e, \end{array} \end{equation} 
from which (\ref{alg in alpha}) gives
\begin{equation}
\begin{array}{rl}
\left[p_a,p_b\right] &=0, \quad [M,N_1]=-N_2, \quad [M,N_2]=N_1, \quad [N_1,N_2]=M,\\[1.0ex]
\left[p_0,M\right]&=0, \quad \left[p_1,M\right]=-p_2 e^{-\frac{p_1}{\sqrt{\lambda}}},\quad \left[p_2,M\right]=e^{-\frac{p_1}{\sqrt{\lambda}}}\left(\frac{ \sqrt{\lambda}}{ 2}(e^{\frac{2p_1}{\sqrt{\lambda}}}-1)- \frac{1}{ 2\sqrt{\lambda}}(p_2^2-p_0^2)\right) \\[1.0ex]
\left[p_0, N_1\right]&=-p_2, \quad [p_0, N_2]=e^{-\frac{p_1}{\sqrt{\lambda}}}\left(\frac{ \sqrt{\lambda}}{ 2}(e^{\frac{2p_1}{\sqrt{\lambda}}}-1)- \frac{1}{ 2\sqrt{\lambda}}(p_2^2-p_0^2)\right), \\[1.0ex]
\left[p_1,N_1\right]&=[p_2,N_2]=0,\quad [p_1,N_2]=p_0 e^{-\frac{p_2}{\sqrt{\lambda}}}, \quad [p_2,N_1]=-p_0. \\[1.5ex]
\end{array} \end{equation}
The co-products are 
 \begin{equation}
\begin{array}{rl}
\Delta p_0 &=p_0\otimes1+e^{\frac{p_1}{\sqrt{\lambda}}}\otimes p_0,\\
\Delta p_1 &=p_1\otimes1+1\otimes p_1,\\
\Delta p_2 &=p_2\otimes1+e^{\frac{p_1}{\sqrt{\lambda}}}\otimes p_2,\\
\Delta M&=1\otimes M+M \otimes e^{-\frac{p_1}{\sqrt{\lambda}}}- \frac{1}{ \sqrt{\lambda}}N_1 \tens p_0 e^{-\frac{p_1}{\sqrt{\lambda}}},\\
\Delta N_1&=1\otimes N_1+N_1\otimes 1,\\
\Delta N_2&=1\otimes N_2+N_2 \otimes e^{-\frac{p_1}{\sqrt{\lambda}}}- \frac{1}{ \sqrt{\lambda}} N_1 \tens p_2 e^{-\frac{p_1}{\sqrt{\lambda}}} . \end{array}  \end{equation}

\subsection{The case $\lambda =0$}

For the case $\lambda=0$ and $c^2>0$ (Lorentzian signature),  the bicrossproduct   in Table~\ref{resulttable} $\C(AN(2))\lrbicross_l U(\mathfrak{sl}(2,\R))$
($\kappa$-Poincar\'e algebra with lightlike deformation parameter).
The translation generators $p_-,p_+,\tilde{p}$ null rotations $N_-,N_+$ and bost $\tilde{M}$  are
\begin{equation}
\begin{array}{rl}
2\alpha &=p_+ ,\quad \xi=\tilde{p}, \quad \eta=p_-, \\
\mbox{and } \tilde{M} &=\frac{1}{2}\tilde{ \bv m} \cdott \bv e,\quad N_-=\frac{ 1}{ 2} \bv n\cdott \bv e, \quad  N_+=\frac{1}{ 2}\tilde{\bv n }\cdott \bv e. \end{array}  \end{equation} 
This gives the algebra
\begin{equation}
\begin{array}{rl}
\left[\tilde{p},p_-\right] &=[\tilde{p},p_+]=[p_-,p_+]=0,\\[1.0ex]
[\tilde{M},N_-] &=N_-, \quad [\tilde{M},N_+]=-N_+, \quad [N_-,N_+]=-\tilde{M},\\[1.0ex]
\left[\tilde{p},\tilde{M}\right]&=0 \quad \left[p_-,\tilde{M}\right]=( p_-+\frac{1}{2}\tilde{p}^2)e^{-p_+},\quad [p_+,\tilde{M}]=e^{-p_+}-1, \\[1.0ex]
\left[\tilde{p}, N\_\right]&=1-e^{p_+}, \quad [\tilde{p}, N_+]=( p_-+\frac{1}{2}\tilde{p}^2)e^{-p_+}, \\[1.0ex]
\left[p_-,N\_\right]&=-\tilde{p}, \quad [p_-,N_+]=0, \quad [p_+,N_-]=0, \quad [p_+,N_+]=\tilde{p} e^{-p_+}.  \\
\end{array} \end{equation}
The co-products are 
 \begin{equation}
\begin{array}{rl}
\Delta \tilde{p} &=\tilde{p}\otimes1+e^{p_+}\otimes \tilde{p},\\
\Delta p_- &=p_-\otimes1+e^{p_+}\otimes p_-,\\
\Delta p_+ &=p_+\otimes1+1\otimes p_+,\\
\Delta \tilde{M} &=1\otimes \tilde{M}-\tilde{M} \otimes e^{-p_+}- N_-\tens \tilde{p} e^{-p_+} ,\\
\Delta N_-&=1\otimes N_-+N_-\otimes 1,\\
\Delta N_+&=1\otimes N_++N_+ \otimes e^{-p_+}+ N_- \tens p_-e^{-p_+} .
\end{array}  \end{equation}

\end{document}